\documentclass[a4paper,onecolumn,11pt,accepted=2022-09-02]{quantumarticle}
\pdfoutput=1

\usepackage[utf8]{inputenc}
\usepackage{amsmath, amssymb, amsthm}
\usepackage{graphicx}
\usepackage{xcolor}
\usepackage{enumitem}
\usepackage{titlesec}
\usepackage{datetime}
\usepackage{hyperref}
\usepackage{import}
\usepackage{mathtools}

\definecolor{crimson}{RGB}{186,0,44}

\theoremstyle{definition}
\newtheorem{definition}{Definition}[section]
\newtheorem{lemma}{Lemma}[section]
\newtheorem{theorem}{Theorem}[section]
\newtheorem{corollary}{Corollary}[theorem]
\newtheorem*{theorem*}{Theorem}
\newtheorem*{corollary*}{Corollary}
\newtheorem{remark}{Remark}[section]
\newtheorem{conjecture}{Conjecture}[section]

\newtheorem{problem}{Problem}[section]

\makeatletter
\newcommand{\subalign}[1]{%
  \vcenter{%
    \Let@ \restore@math@cr \default@tag
    \baselineskip\fontdimen10 \scriptfont\tw@
    \advance\baselineskip\fontdimen12 \scriptfont\tw@
    \lineskip\thr@@\fontdimen8 \scriptfont\thr@@
    \lineskiplimit\lineskip
    \ialign{\hfil$\m@th\scriptstyle##$&$\m@th\scriptstyle{}##$\hfil\crcr
      #1\crcr
    }%
  }%
}
\makeatother

\begin{document}

\title{Multivariable quantum signal processing (M-QSP): prophecies of the two-headed oracle}
\author{Zane M.\ Rossi}\affiliation{%
Department of Physics, Massachusetts Institute of Technology, Cambridge, Massachusetts 02139, USA}
\orcid{0000-0002-7718-654X}
\author{Isaac L.\ Chuang}
\affiliation{%
Department of Physics, Department of Electrical Engineering and Computer Science, and Co-Design Center for Quantum Advantage,
Massachusetts Institute of Technology, Cambridge, Massachusetts 02139, USA}

\begin{abstract}

\noindent Recent work shows that quantum signal processing (QSP) and its multi-qubit lifted version, quantum singular value transformation (QSVT), unify and improve the presentation of most quantum algorithms. QSP/QSVT characterize the ability, by alternating ansätze, to obliviously transform the singular values of subsystems of unitary matrices by polynomial functions; these algorithms are numerically stable and analytically well-understood. That said, QSP/QSVT require consistent access to a \emph{single} oracle, saying nothing about computing \emph{joint properties} of two or more oracles; these can be far cheaper to determine given an ability to pit oracles against one another coherently.
    
This work introduces a corresponding theory of QSP over multiple variables: M-QSP. Surprisingly, despite the non-existence of the fundamental theorem of algebra for multivariable polynomials, there exist necessary and sufficient conditions under which a desired \emph{stable} multivariable polynomial transformation is possible. Moreover, the classical subroutines used by QSP protocols survive in the multivariable setting for non-obvious reasons, and remain numerically stable and efficient. Up to a well-defined conjecture, we give proof that the family of achievable multivariable transforms is as loosely constrained as could be expected. The unique ability of M-QSP to \emph{obliviously} approximate \emph{joint functions} of multiple variables coherently leads to novel speedups incommensurate with those of other quantum algorithms, and provides a bridge from quantum algorithms to algebraic geometry.

\end{abstract}

\maketitle

\section{Introduction}

    Recent advances in the theory of quantum algorithms have led to a powerful method, quantum singular value transformation (QSVT), for applying polynomial transformations to the singular values of sub-blocks of unitary processes \cite{gslw_19}. These algorithms demonstrate the ability of simple circuit ansätze to expressively and efficiently control quantum subsystem dynamics. In addition to improving the performance of many known quantum algorithms \cite{rall_21, coherent_ham_sim_21}, QSVT has great explanatory utility: unifying the presentation of most major known quantum algorithms \cite{mrtc_21}. This includes Hamiltonian simulation \cite{lloyd_hamiltonian_qsvt_21, coherent_ham_sim_21}, search \cite{gslw_19}, phase estimation \cite{rall_21}, quantum walks \cite{gslw_19}, fidelity estimation \cite{gilyen_fidelity_22}, sophisticated techniques for measurement \cite{petz_recovery_20}, and channel discrimination \cite{rossi_qht_21}. QSVT has found purchase in surprisingly disparate subfields, from undergirding a general theory of quantum-inspired classical algorithms for low-rank machine learning \cite{chia_20}, to quantum cryptographic protocols with zero-knowledge properties \cite{lombardi_21}.
    
    At its core, QSVT is a sophisticated, lifted, multi-qubit extension to quantum signal processing (QSP) \cite{lyc_16_equiangular_gates, lc_17_simultation, lc_19_qubitization}, which itself completely characterizes the achievable polynomial transformations of a scalar value embedded in a single-qubit rotation. Again, although the transformations possible with QSP are with respect to a simple circuit ansatz, they are surprisingly general, and it is the simplicity of this ansatz in conjunction with its expressibility that leads to QSP's and QSVT's numerical stability \cite{haah_2019, chao_machine_prec_20, dong_efficient_phases_21} and usefulness. This single-qubit algorithm can be lifted to arbitrarily large systems of qubits through the identification of natural qubit-like subspaces within high-dimensional unitaries, leading to the famed exponential performance improvements of quantum algorithms for, e.g., matrix inversion \cite{hhl_09}, factoring \cite{shor_99}, and simulation \cite{feynman_18}.
    
\begin{figure}[htpb]
        \centering
        \includegraphics[width=0.9\textwidth]{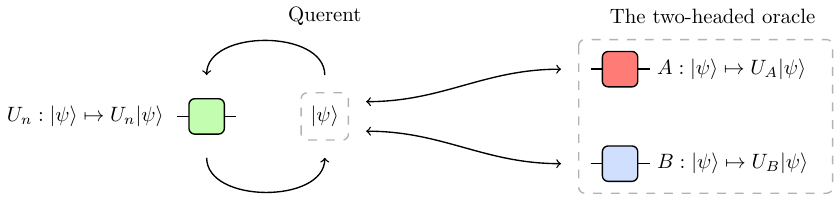}
        \caption{A meeting between the intrepid querent and the two-headed quantum oracle. The querent can hold a quantum state and may submit it to the unitary action of \emph{either} head, $A$ or $B$, whose actions are consistent. The querent may submit again and again to either oracle (possibly a different one each time) with this same state, and may intersperse their own unitary gates $U_n$ along the way before finally measuring. The querent seeks, of course, to interpret the oracle's fragile mystery.}
        \label{fig:m_qsp_sibyl_diagram}
    \end{figure}

    A natural extension to QSP considers the scenario when a computing party is given access to not just one oracle encoding a scalar signal, but two such oracles whose relation is unknown in general, as in Fig.~\ref{fig:m_qsp_sibyl_diagram}. One can view this setting as a game, or else a coherent interrogation between a quantum querent and a novel \emph{two-headed} unitary oracle. As in standard QSP, the querent hopes to decide on hidden properties of the oracle(s). The motivation for this extension stems from a general interest in inference/communication: if one is interested only in \emph{joint} properties of two signals, rather than \emph{absolute} properties, do there exist corresponding realizable efficiencies in methods to decide on these properties? Moreover, can these transformations be done entirely \emph{coherently}, as is one major benefit of QSP/QSVT? Any setting wherein one hopes to subject a quantum system to multiple interleaved unitary operations, and equivalently where one hopes to talk of multivariable polynomial transformations (of general interest in classical and quantum computer science both \cite{ding_m_pke_09, aaronson_q_money_12, anshu_21}), promotes investigation into multivariable analogues of QSP/QSVT.
    
    These considerations, in conjunction with inspiration by the success of standard QSP, suggest the need for a theory of multivariable polynomial transforms embedded in unitary matrices. The theory of multivariable polynomials, even outside the scope of quantum information, is substantively more complex for fundamental reasons in algebraic geometry including the loss of the fundamental theorem of algebra, and thus provides a highly non-trivial elaboration on the theory of QSP-like algorithms. The fruit of this effort is worthwhile, though, and provides additional methods for showing separations between the performance of quantum algorithms for inference and their classical counterparts, as well as an opportunity to leverage the deeply-studied mathematical subfield of algebraic geometry in many dimensions for purposes of practical computing interest.
    
    This work makes progress toward a comprehensive theory of achievable multivariable polynomial transformations for QSP-like ansätze, constrained mostly to the two-variable setting. In generic terms, as opposed to standard QSP which is the study of achievable functions from the circle to SU(2), M-QSP considers functions from the multitorus to SU(2)
        \begin{equation}
            \underset{\text{Standard QSP}}{\mathbb{T} \rightarrow \text{SU(2)}}
            \quad\quad\quad    
            \underset{\text{M-QSP}}{\mathbb{T}^n \rightarrow \text{SU(2)}},
        \end{equation}
    where $\mathbb{T}$ is the set of complex numbers of modulus one. We will primarily study the case $n = 2$. The business of QSP \cite{haah_2019, lyc_16_equiangular_gates, lc_17_simultation, lc_19_qubitization, gslw_19} and M-QSP comes in two fundamental directions: (1) describing how the parameters $\Phi$ defining a circuit ansatz are taken to polynomial transforms (the so-called $\Phi \mapsto P, Q$ direction), and (2) providing simple conditions under which a suitable \emph{partial specification} of polynomial transforms $\tilde{P}, \tilde{Q}$ (still useful for solving a desired algorithmic problem) can be `completed' and their corresponding $\Phi$ calculated (the so-called $\tilde{P}, \tilde{Q} \mapsto \Phi$ direction). These directions are made explicit in Sec.~\ref{sec:qsp_constructions}.
    
\begin{figure}[htpb]
        \centering
        \includegraphics[width=1.0\textwidth]{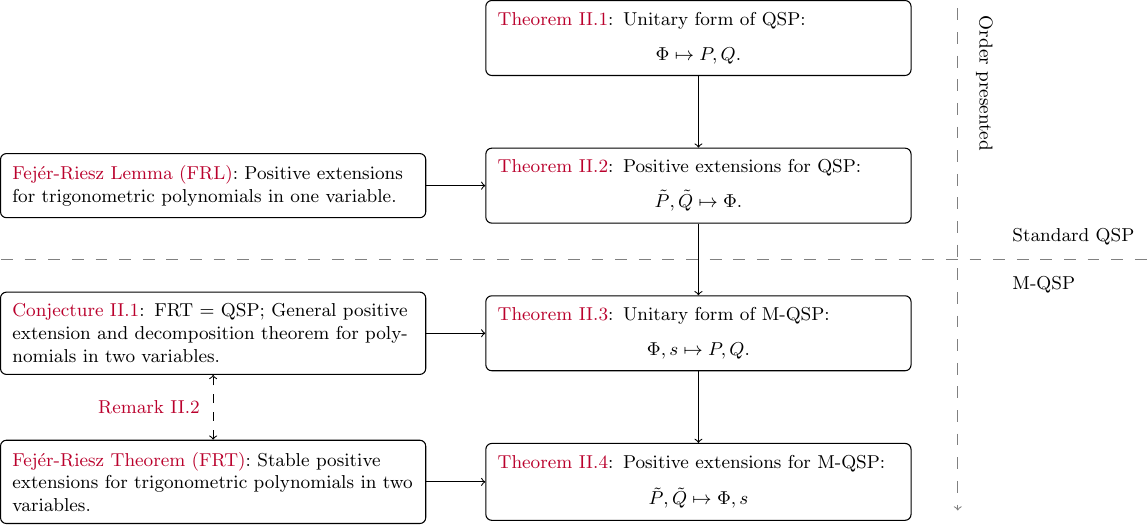}
        \caption{A summary of the major theorems and related conjectures of this work. Within standard QSP (top half), the theorems are given without qualification; Thm.~\ref{thm:partial_qsp_properties} is shown to depend entirely on the single-variable Fejér-Riesz lemma. Crossing the horizontal dotted line to M-QSP (bottom half), we give parallel theorems, save Thm.~\ref{thm:major_m_qsp_properties} now depends on Conjecture~\ref{conj:m_qsp_factorability}, and Thm.~\ref{thm:partial_m_qsp_properties}, now depends on the intricate multivariable Fejér-Riesz theorem (FRT), itself reliant on Conjecture~\ref{conj:m_qsp_factorability} to yield valid M-QSP phases for a desired multivariable polynomial transformation. The relation between the FRT and Conjecture~\ref{conj:m_qsp_factorability} (the FRT = QSP conjecture) is discussed in Remark~\ref{rem:conjecture_frt_relation}.}
        \label{fig:conjecture_flow_diagram}
    \end{figure}

    Study of such functions relies on powerful extensions of results from single-variable functional analysis and approximation theory to multivariable settings. To keep this elaboration organized, we introduce theorems of standard QSP and M-QSP in parallel, taking care to state assumptions, methods, and underlying theorems that distinguish these two paths. The loss of foundational theorems in the multivariable setting means M-QSP theorems are more strongly qualified. This structure is presented diagrammatically in Figure~\ref{fig:conjecture_flow_diagram}, and discussed in Sec.~\ref{sec:paper_outline}.
    
\subsection{Prior work}

    As this work proposes a substantively distinct algorithm drawing on methods/assumptions common in the study of QSP, we give a roadmap to prior major work. This work can be laid along three broad directions: (1) fundamental work on the analytic form of of QSP/QSVT protocols, (2) detailed work on the stability of numerically implementing optimizations over, and the classical subroutines of, QSP/QSVT and (3) concrete applications of QSP/QSVT as a `meta-algorithm' or `algorithmic framework' to previously unconsidered problems.
    
    (1) The theory of QSP has its origin in the study of composite pulse techniques in NMR \cite{lyc_16_equiangular_gates}, though its first instantiation by name appeared in service of improved methods for Hamiltonian simulation \cite{lc_17_simultation, lc_19_qubitization}. These papers fleshed out theorems on the structure and numerical stability of QSP. The curious ability to locate invariant qubit-like subspaces in larger Hilbert spaces and perform QSP simultaneously within them, \emph{obliviously to the eigenbases or singular vector bases of these subsystems as well as the eigenvalues or singular values}, led to the far expanded QSVT \cite{gslw_19}, whose uses, robustness, and applications \cite{mrtc_21, gslw_19} have been recently explored. Finally, rephrasing QSVT in terms of Hamiltonian simulation \cite{lloyd_hamiltonian_qsvt_21} has both simplified the presentation and in some ways brought this algorithmic story full circle. Ongoing work continues to simplify the presentation of these algorithms.
    
    (2) Notably, initially proposed constructive methods for determining the defining parameters $\Phi$ of a QSP ansatz were known to be numerically unstable. Surprisingly, extensive results have since shown that there exist novel, stable, divide-and-conquer classical methods for determining these parameters \cite{chao_machine_prec_20, dong_efficient_phases_21, haah_2019}. In addition to standard approximative constraints for the embedded polynomial transformations, new work has also investigated more sophisticated constraints and their effect on QSP performance \cite{sarkar_density_22}. Moreover, beyond algorithms for polynomial approximation and QSP phase read-off, recent forays into symmetrized (restricted) QSP ansätze \cite{sym_qsp_21} have proven that the relevant loss landscapes are, under some reasonable restrictions, convex.
    
    (3) Finally, QSP/QSVT have recently been applied to a wide variety of subfields, both in and out of quantum information. These include Hamiltonian simulation \cite{coherent_ham_sim_21}, phase estimation \cite{rall_21}, quantum zero-knowledge proofs \cite{lombardi_21}, classical quantum-inspired machine learning algorithms \cite{chia_20}, semi-definite programming \cite{q_sdp_solvers_20}, quantum adiabatic methods \cite{lin_eig_filter_20}, the approximation of correlation functions \cite{rall_correlation_20}, the approximation of fidelity \cite{gilyen_fidelity_22}, recovery maps \cite{petz_recovery_20}, and fast inversion of linear systems \cite{tong_inversion_21}. Efforts continue to bring further computational problems into the fold of QSP/QSVT.
    
    This work, by merit of considering a distinct ansatz that precludes many of the proof methods of standard QSP, is somewhat incommensurate with the prior work given. However, there is good reason to believe that much of the wonderful studies into the numerical stability of standard QSP may be (with suitable modification) applicable in the multivariable setting. Moreover, we hope that a variety of new, previously unconsidered problems from the classical world may now fall under the purview of M-QSP, drawing on its use of theorems originally designed for understanding autoregressive filter design and image analysis \cite{mv_frt_04}. In a sense this work advocates going back to the \emph{bare metal} of QSP algorithms: reconsidering its basic ansatz, elaborating on tweaks to its setting, and demarcating a family of \emph{QSP-like} ansätze toward a better understanding of quantum algorithms.

\subsection{Paper outline and informal theorem statements} \label{sec:paper_outline}

    This work introduces multivariable QSP (M-QSP) in the terms of QSP (and in parallel with it), and thus assumes some familiarity with the constitutive theorems of the latter. Concretely, Sec.~\ref{sec:qsp_review} introduces two fundamental theorems of QSP that are argued to be representative of the statements one would like to be able to make to pragmatically understand its use. Next, Sec.~\ref{sec:alg_geo_review} briefly covers concepts in algebraic geometry that will support the major theorems of M-QSP in Sec.~\ref{sec:m_qsp_review}. The theorems of Sec.~\ref{sec:m_qsp_review}, given casual statements below, are modelled quite closely after those of standard QSP, but require far more involved techniques to resolve, as shown in Appendix~\ref{appx:m_qsp_proofs}. Here, we give the informal statements of the major results of this work. Depiction of their constitutive assumptions and relations is given in Figure~\ref{fig:conjecture_flow_diagram}.

    \begin{theorem*}
        \emph{Unitary form of M-QSP.} Informal statement of Theorem~\ref{thm:major_m_qsp_properties}. All M-QSP protocols, suitably defined by an alternating ansatz (Def.~\ref{def:m_qsp}), have a simple unitary form with parity, norm, and determinant constraints, up to the resolution of Conjecture~\ref{conj:m_qsp_factorability} (here named the FRT = QSP conjecture).
    \end{theorem*}
    
    \begin{theorem*}
        \emph{Positive extensions for M-QSP.} Informal statement of Theorem~\ref{thm:partial_m_qsp_properties}. Given a desired multivariable polynomial transform that satisfies parity and norm constraints simpler than those in Theorem~\ref{thm:major_m_qsp_properties}, the ability to find an M-QSP protocol that embeds this transform depends solely on whether a well-defined matrix of Fourier components of a related transform has low rank, under the assumption the related transform is \emph{stable} (Def.~\ref{def:stable_polynomials}).
    \end{theorem*}
    
    \begin{corollary*}
        \emph{Uniqueness of stable positive extensions for M-QSP.} Informal statement of Corollary~\ref{cor:m_qsp_phase_readoff}. Given a unitary matrix of a valid M-QSP protocol, the real numbers parameterizing the M-QSP protocol can be determined by an efficient classical algorithm.
    \end{corollary*}
    
    In addition to showing presenting basic theorems M-QSP up to certain assumptions, we conclude this work by providing a few worked examples in Sec.~\ref{sec:worked_examples}, explicitly providing decision problems (here in a noiseless setting) for which M-QSP provides an intuitive and quantitatively easy to show improvement over other algorithms. We present problems for which the approach of M-QSP is natural, and for which no other obvious quantum algorithmic methods are known. In essence, we show the ability for M-QSP protocols to decide on \emph{joint properties} of pairs of oracles, where these properties cannot be determined as efficiently (or even deterministically) by serial or parallel standard QSP protocols using each oracle individually; in other words, there exist scenarios where it is far better to compute functions coherently `under the hood' of a unitary evolution, followed by a precise measurement, as opposed to classically following an estimative process. We show that there exist non-trivial examples of algorithmic advantage in query complexity for M-QSP protocols which permit no efficient reductions to a single-variable settings (Problem~\ref{prob:multi_channel_discrimination}), and provide accompanying geometric intuitions for why this is the case.
    
    Finally, and more abstractly, taking inspiration from standard QSP in its natural connection to the famed Chebyshev polynomials, we use M-QSP to define one infinite family of multivariable Chebyshev-like polynomials and discuss their significance. In turn, we point toward the usefulness of M-QSP in studying the theory of orthogonal polynomials, which has great relevance to the theory of positive extensions and signal processing (both classical and quantum). 
    
    For the interested reader, discussion on the outlook for M-QSP, its caveats, its open problems, and its position in the pantheon of quantum algorithms, is presented in Sec.~\ref{sec:discussion}.

\section{Construction and analysis of M-QSP} \label{sec:qsp_constructions}

    This section has two goals: (1) a quick re-introduction of the major theorems (and moral takeaways) of QSP drawn from \cite{gslw_19} (with some alternative proofs), followed by (2) a series of appropriate definitions, lemmas, and finally analogous theorems for M-QSP. Proofs, applications, caveats, and worked examples, are left to Secs.~\ref{sec:discussion}, \ref{sec:worked_examples}, and Appendix~\ref{appx:m_qsp_proofs}.

    \subsection{Review of standard (single-variable) QSP} \label{sec:qsp_review}
    
        In standard QSP a computing party is given oracle access to a unitary operation $A(x) = \exp{(i\arccos{x}\,\sigma_x)}$ for some unknown $x \in [-1,1]$; this oracle performs a consistent, unknown rotation about a known axis on the Bloch sphere. By repeatedly applying this oracle, interspersed with known rotations about a separate orthogonal (in this case $\sigma_z$) axis, the unknown signal can be cleverly correlated with its previous applications to generate complex polynomial functions of the parameter $x$. This transformation can be done \emph{oblivious} to $x$. The circuit for QSP, as well as its defining parameters, are depicted in Figure~\ref{fig:m_qsp_specification}. Characterizing the possible polynomial transforms is the business of QSP, and leads to surprising applications and performance improvements for many known quantum algorithms as discussed previously. We quote two important theorems; the first (Theorem~\ref{thm:major_qsp_properties}) characterizes the explicit unitary form of the QSP ansatz, while the second (Theorem~\ref{thm:partial_qsp_properties}) introduces a simple necessary and sufficient condition (and an implicit constructive method) under which a desired polynomial transform can be embedded. We refer readers to the original proofs of these theorems, and provide an alternative proof for Theorem~\ref{thm:partial_qsp_properties} in Appendix~\ref{appx:m_qsp_proofs}.
        
        \begin{theorem} \label{thm:major_qsp_properties}
            Unitary form of quantum signal processing (QSP), i.e., $\Phi \mapsto P, Q$. Theorem 3 in \cite{gslw_19}. Let $n \in \mathbb{N}$. There exists $\Phi = \{\phi_0, \phi_1, \cdots, \phi_n\} \in \mathbb{R}^{n + 1}$ such that for all $x \in [-1, 1]$:
                \begin{equation} \label{eq:qsp_def}
                    U_{\Phi}(x) 
                    = 
                    e^{i\phi_0\sigma_z}\prod_{k = 1}^{n} \left(A(x)\, e^{i\phi_k \sigma_z}\right)
                    =
                    \begin{pmatrix}
                        P(x) & i\sqrt{1 - x^2}Q(x)\\
                        i\sqrt{1 - x^2}Q^*(x) & P^*(x)
                    \end{pmatrix},
                \end{equation}
            where $A(x) = \exp{(i\arccos{x}\,\sigma_x)}$ if and only if $P, Q \in \mathbb{C}[x]$ such that
                \begin{enumerate}[label=(\arabic*)]
                    \item $\text{deg}(P) \leq n$ and $\text{deg}(Q) \leq n - 1$.
                    \item $P$ has parity-$n \pmod 2$ and $Q$ has parity-$(n-1)\pmod 2$.
                    \item For all $x \in [-1, 1]$ the relation $|P(x)|^2 + (1 - x^2)|Q(x)|^2 = 1$ holds.
                \end{enumerate}
            Note here and elsewhere $\sigma_x, \sigma_z$ are the usual single-qubit Pauli matrices with non-zero entries $\pm 1$.
        \end{theorem}
        
        \begin{theorem} \label{thm:partial_qsp_properties}
            Reconstruction of QSP protocols from partial embeddings, i.e., $\tilde{P}, \tilde{Q} \mapsto \Phi$. Theorem 5 in \cite{gslw_19}. Let $n \in \mathbb{N}$ fixed. Let $\tilde{P}, \tilde{Q} \in \mathbb{R}[x]$. There exists some $P, Q \in \mathbb{C}[x]$ satisfying conditions (1-3) of Theorem~\ref{thm:major_qsp_properties} such that $\tilde{P} = \Re(P)$ and $\tilde{Q} = \Re(Q)$ if and only if $\tilde{P}, \tilde{Q}$ satisfy conditions (1-2) of Theorem~\ref{thm:major_qsp_properties} and for all $x \in [-1, 1]$
                \begin{equation}
                    |\tilde{P}(x)|^2 + (1 - x^2)|\tilde{Q}(x)|^2 \leq 1.
                \end{equation}
            The same holds if we replace real parts by imaginary parts and additionally $\tilde{P} \equiv 0$ or $\tilde{Q} \equiv 0$ can be chosen for simplicity. An alternative proof to that of \cite{gslw_19} is given in Appendix \ref{appx:m_qsp_proofs}, relying on the single-variable Fejér-Riesz lemma, working in the Laurent picture. A similar method is used in the multivariable case.
        \end{theorem}
        
        The two theorems above were selected out of \emph{many} given in the major references \cite{gslw_19, lyc_16_equiangular_gates, lc_17_simultation, lc_19_qubitization}. These together give a minimal toolkit for using and thinking about QSP. Theorem~\ref{thm:major_qsp_properties} clearly states what QSP unitaries must look like (i.e., going from real phases to embedded polynomials), while Theorem~\ref{thm:partial_qsp_properties} addresses the reverse problem (going from desired embedded polynomials to real phases). Moreover, Theorem~\ref{thm:partial_qsp_properties} is a problem in matrix completions: given a \emph{partially specified} QSP unitary whose elements are simply constrained, can one find missing components that satsify a more complex constraint? Consequently, for reasons to be discussed later, this completed matrix then immediately yields a set of QSP phases $\Phi$.
        
        The utility of matrix completions is clear; note that in Theorem~\ref{thm:partial_qsp_properties} one can choose $\tilde{Q} = 0$ identically, and observe that the $\lvert + \rangle \mapsto \lvert +\rangle$ transition probability is simply $|\tilde{P}(x)|^2$. The polynomial transformation is directly accessible for sampling. There are of course many ways to partially specify a QSP protocol (discussed in prior work), but we claim that Theorem~\ref{thm:partial_qsp_properties} essentially captures their foundation. Consequently, the alternative proof in the appendix provides a concrete connection from the theory of matrix completions to methods vital to understanding the multivariable setting.
        
        As a final note, we want to emphasize again that QSP performs its transformations \emph{obliviously}, that is, independent of $x$. By modifying the functional forms approximated by the embedded polynomials, important properties of $x$ can often be computed more cheaply than if one were first to estimate $x$ and apply a classical computation to this classical result; moreover, because the circuit is coherent, the result of this transformation can be used for further quantum computation. In the two variable case, this oblivious transformation becomes even more useful; i.e., one may only want to compute a joint property of two variables, deciding on properties of their correlation, rather than each variable itself. Maintaining coherence and obliviousness of these functional transforms, and thus avoiding classical post-processing at all costs, is essential to the great algorithmic savings possible with QSP/QSVT.
        
    \subsection{QSP and algebraic geometry} \label{sec:alg_geo_review}
    
        We give a casual map for the intrepid but non-specialist reader to results in algebraic geometry that permit the proofs of Theorems~\ref{thm:major_m_qsp_properties} and \ref{thm:partial_m_qsp_properties} in Appendix~\ref{appx:m_qsp_proofs}. It is the hope that merely introducing some common terms and ideas may lead others to the application of similar techniques to low-hanging fruit in the theory of QSP-like algorithms.
        
        The utility of QSP rests in the ability of the computing party to choose a simply-constrained polynomial transform and efficiently determine QSP phases achieving this transform. It turns out that this ability to simply specify a desired polynomial (see proof of Theorem~\ref{thm:partial_qsp_properties}) relies on a special fact about positive trigonometric polynomials. More specifically, these results reside in a family of what are known as \emph{positivstellensätze} (positive-locus-theorems), or for non-negative polynomials \emph{nichtnegativstellensätze} (nonnegative locus-theorems). In turn these belong to an even larger family of \emph{nullstellensäztze} (zero-locus-theorems). These families of theorems seek to establish a relationship between geometry and algebra going back to Diophantus (and more recently Hilbert, Weil, Riemann, Grothendieck and Gröbner, to name but a few). Here are a few examples of the types of statements common to these subfields \cite{marshall_sos_08}.
            \begin{enumerate}[label=(\arabic*)]
                \item All positive single-variable polynomials over the reals can be written as the sum of at most two squares.
                \item All multivariate polynomials that take only nonnegative values over the reals can be written as sums of squares of real rational functions. (Hilbert's 17\textsuperscript{th} problem.)
                \item All nonnegative trigonometric polynomials can be written as exactly one square.
            \end{enumerate}
        These theorems seek to establish succinct descriptions of algebraic functions (often polynomials, for our purposes) which satisfy constraints on algebraic and semialgebraic sets (often products of intervals, for our purposes) \cite{dumitrescu_monograph_07}. In general these statements get weaker as the number of variables increases (see Hilbert's 17\textsuperscript{th} problem above) \cite{dritschel_outer_05, geronimo_ortho_poly_07, mv_frt_04}, and stronger as the family of considered polynomials is restricted (see the Fejér-Riesz lemma, Lemma~\ref{lemma:frt}). Moreover, such results can also often be extended to polynomials over operators as well \cite{op_val_frt_09}.
        
        \begin{remark}
            Here we provide a rapid series of quick, casual definitions of common terms in this work. Algebraic sets are subsets of (for us) $\mathbb{R}^n$ or $\mathbb{C}^n$ defined by zeros of finite sets of polynomial expressions. Sometimes it is required or preferred that one discusses only irreducible algebraic sets, referred to as algebraic varieties (and where algebraic sets can be thought of as finite unions of algebraic varieties). Semi-algebraic sets are defined the same way, save as the locus of both roots of finitely many polynomial equations and the solutions to finitely many polynomial inequalities. They are also preserved under finite union and intersection. In principle these sets and varieties can be defined over general algebraically closed fields, i.e., fields $F$ where every non-constant polynomial in $F[x]$ (univariate polynomials with coefficients in $F$) has a root in $F$. While the definitions presented here can be made extremely general (e.g., the irreducible algebraic sets above can be defined purely in terms of prime ideals of polynomial rings and closed subsets in the Zariski topology), we try to keep close to $\mathbb{R}^n$ and $\mathbb{C}^n$.
            
            A foundational observation of algebraic geometry is the formalization of the often overlooked fact that a univariate polynomial (an algebraic object) is uniquely defined by its root-set in the corresponding field (a geometric object). For more general fields or in multivariable settings, similar statements connect ideals of polynomial rings\footnote{We don't define these here because they won't be needed for the proofs given, but they are vital, simple building blocks of the statements of algebraic geometry and abstract algebra generally. For some classic textbooks see \cite{hartshorne_textbook} and the more introductory \cite{fulton_textbook}.} and algebraic sets (e.g., Hilbert's Nullstellensatz). By understanding algebraic objects through geometric means (and vice versa) a variety of otherwise difficult problems can become amenable to simpler, otherwise established methods of analysis.
        \end{remark}
        
        This work will study a subset of Laurent polynomials, trigonometric polynomials, for which relevant positivstellensätze are known to exist for two variables, albeit under complicated restrictions \cite{mv_frt_04, dumitrescu_monograph_07}. The primary work of Theorem~\ref{thm:partial_m_qsp_properties} will be thus to guarantee that the setting of M-QSP satisfies the requirements for the application of these positivstellensätze, and to probe the induced properties of this application. We aim also to interpret the meaning of these rather arcane methods in the context of quantum algorithms.
        
        This work mostly relies on the generalization of a standard result for non-negative single-variable trigonometric polynomials known and attributed variously \cite{polya_szego_analysis} as the Fejér-Riesz lemma (or theorem), given in Lemma~\ref{lemma:frt}. This lemma states that such polynomials can be reëxpressed as squares uniquely up to simple constraints. We claim, and show in the alternative proof of \ref{thm:partial_qsp_properties}, that this lemma is effectively the only non-trivial mathematical fact made use of in QSP. The purpose of this lemma in both QSP and M-QSP is to enable a partial specification of a unitary matrix to be completed under constraints. This completed (i.e., filled-out) unitary is easily implementable as a series of simple quantum gates. That is, if these partial specifications can be satisfied, then the other major theorem of QSP (Theorem~\ref{thm:major_qsp_properties}) guarantees that the corresponding QSP phases can be determined efficiently (where further work has shown numerical stability as well).
        
        In the multivariable setting, a generalized Fejér-Riesz theorem (now truly a theorem, with proof in excess of thirty pages \cite{mv_frt_04}) exists! We make use of this theorem to undergird a new theory of M-QSP, and invite other researchers to make use of the rich related literature to attack similar problems in the theory of alternating quantum circuit ansätze.
        
        \begin{remark}
            Note that the Fejér-Riesz theorem is disjoint from the classical algorithms used to find the QSP phases of a QSP protocol given its corresponding unitary, as well as classical optimization algorithms used to find good polynomial approximations to continuous functions (e.g., the Remez-type exchange and Parks-McClellan algorithms \cite{grenez, remez, parks_mcclellan_73}). Both of these classical algorithmic families require far fewer modifications when moving to the multivariable case than the Fejér-Riesz theorem.
        \end{remark}
    
    \subsection{Extension to M-QSP} \label{sec:m_qsp_review}
    
        We introduce a series of definitions toward analogues of Theorems~\ref{thm:major_qsp_properties} and \ref{thm:partial_qsp_properties} under the assumption of oracle access to two signal operators whose relation is unknown in general. This involves two steps: (1) a simple transformation of variables to clarify the statement of the major theorems, and (2) an application of a series of results from the theory of multivariable nullstellensätze. We start with the definition of multivariable quantum signal processing (M-QSP) protocols, solidifying the intuition given by Fig.~\ref{fig:m_qsp_sibyl_diagram}.

\begin{figure}[htpb]
                \centering
                \includegraphics[width=1.0\textwidth]{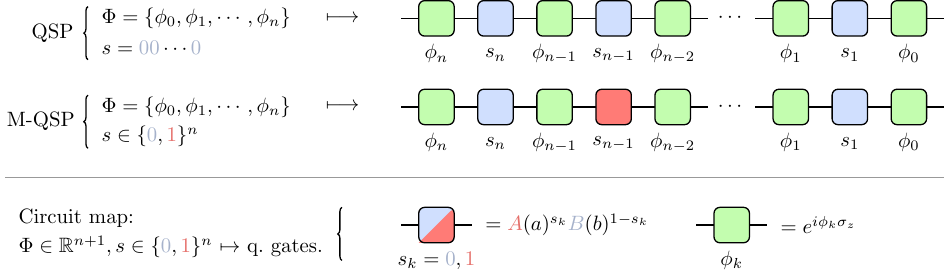}
                \caption{Circuit definitions for standard QSP and M-QSP, indicating the explicit map from the set of real phases $\Phi$ (the QSP phases), and the bit-string $s$ to single-qubit quantum circuits comprising $X$ and $Z$ rotations. This realizes the generic protocol set forth in Fig.~\ref{fig:m_qsp_sibyl_diagram}. M-QSP circuits come from an exponentially large family in which the querent is allowed to apply either the unknown $A(a) = e^{i\theta_a \sigma_x}$ or $B(b) = e^{i\theta_b \sigma_x}$ between $Z$ rotations parameterized by $\Phi$. Here $a = e^{i\theta_a}$ and $b = e^{i\theta_b}$. For standard QSP $s = 11\cdots1$ or $s = 00\cdots0$ depending on the variable of choice. Here $A(a)$ versus $A(x_A)$ in Eq.~\ref{def:m_qsp} refer to the Laurent picture (Def.~\ref{def:laurent_picture}) and $x$ picture respectively; likewise for $B$.}
                \label{fig:m_qsp_specification}
            \end{figure}

            \begin{definition} \label{def:m_qsp}
                Multivariable quantum signal processing (M-QSP). Given oracle access to two unitary operators $A(x_A) = \exp{(i\arccos{x_A}\,\sigma_x)}$ and $B(x_B) = \exp{(i\arccos{x_B}\,\sigma_x)}$, an M-QSP protocol of length $n$ is defined by the length-$n$ binary string $s \in \{0, 1\}^n$ and a set of real phases $\Phi = \{\phi_0, \phi_1, \cdots, \phi_n\} \in \mathbb{R}^{n + 1}$ according to the map to the following quantum circuit:
                    \begin{equation} \label{eq:m_qsp_form}
                        U_{(s, \Phi)}(x_A, x_B) = e^{i\phi_0 \sigma_z}\prod_{k = 1}^{n} A^{s_k}(x_A)B^{1 - s_k}(x_B)e^{i\phi_k \sigma_z}.
                    \end{equation}
                This is again a product of rotations about orthogonal axes on the Bloch sphere. This map is depicted in Figure~\ref{fig:m_qsp_specification}.
            \end{definition}
            
        While Def.~\ref{def:m_qsp} shares obvious similarities to the circuit given in Theorem~\ref{thm:major_qsp_properties} (its major novel component being the bit-string $s$ defining the order of $A, B$ iterates), its form is awkward. For the rest of the work we will use a mathematically equivalent picture (Def.~\ref{def:laurent_picture}) to more smoothly apply methods from algebraic geometry. Similar maps have been considered in \cite{haah_2019, mrtc_21} to improve numerical stability of algorithms for finding standard QSP phases.
        
            \begin{definition} \label{def:laurent_picture}
                The Laurent picture. It will be helpful to consider M-QSP under the map $(x_A, x_B) \in [-1,1]^2$ to $(a, b) \in \mathbb{T}^{2} = \{a, b \in \mathbb{C}^2 \;\text{s.t.}\; |a| = |b| = 1\}$ following from replacing
                    \begin{equation}
                        x_A \mapsto \frac{1}{2}(a + a^{-1}), \quad x_B \mapsto \frac{1}{2}(b + b^{-1}).
                    \end{equation}
                Polynomial transforms in the $x$-picture are equivalent to pseudo-polynomial transforms in the Laurent picture. Often however we will drop the pseudo prefix and simply refer to both as polynomial transforms with the requisite caveats.
            \end{definition}
        
        We will often be working with Laurent polynomials that are real on $\mathbb{T}^2$, so-called \emph{Hermitian trigonometric polynomials} which have the generic form (in two variables)
            \begin{equation} \label{eq:hermitian_trig_poly}
                g(a, b) = \sum_{j = -m}^{m}\sum_{k = -n}^{n} g_{j,k} a^j b^k, \quad g_{j,k} = g_{-j, -k}^{*},
            \end{equation}
        for nonnegative integers $m, n$, where the $g_{j,k}$ are complex. We define the degree of a polynomial like that in Eq.~\ref{eq:hermitian_trig_poly} by the ordered tuple $(m, n)$, and say that a polynomial of degree $(p, q)$ satisfies $(p, q) \preccurlyeq (m, n)$ if both $p \leq m$ and $q \leq n$. We say such a polynomial as in Eq.~\ref{eq:hermitian_trig_poly} has inversion parity $(d_A, d_B) \pmod{2}$ if under the map $a \mapsto a^{-1}$ the polynomial transforms as $g \mapsto (-1)^{d_A}g$, and analogously for $b \mapsto b^{-1}$. We say, toward the following lemma, that a multivariable polynomial in the $x$ picture has negation parity $(d_A, d_B) \pmod{2}$ if under the map $x_A \mapsto -x_A$ the polynomial tranaforms as $g(x_A, x_B) \mapsto (-1)^{d_A}g(x_A, x_B)$, and analogously for $x_B \mapsto -x_B$. Note that one can also consider negation parity in the Laurent picture.
        
        Before moving entirely to the Laurent picture we state a more familiar $x$ picture form of the future Theorem~\ref{thm:major_m_qsp_properties} in part to demonstrate the necessity for moving to the Laurent picture.
    
        \begin{lemma} \label{lemma:m_qsp_form_x}
            Unitary form of multivariable quantum signal processing (M-QSP) in the $x$ picture. Let $n \in \mathbb{N}$. There exists $\Phi = \{\phi_0, \phi_1, \cdots, \phi_n\} \in \mathbb{R}^{n + 1}$ such that for all $x_A, x_B \in [-1,1]^2$ the circuit presented of length $n$ in Definition~\ref{def:m_qsp} with $m = |s|$ (the Hamming weight of $s$) has the form:
                \begin{align}
                    U_{s, \Phi}(x_A, x_B)
                    &=\nonumber\\
                    &\begin{pmatrix}
                        P + Q\sqrt{1 - x_a^2}\sqrt{1 - x_b^2} & R\sqrt{1 - x_a^2} + S\sqrt{1 - x_b^2}\\
                        -R^*\sqrt{1 - x_a^2} -S^*\sqrt{1 - x_b^2} & P^* + Q^*\sqrt{1 - x_a^2}\sqrt{1 - x_b^2}
                    \end{pmatrix},
                \end{align}
            if and only if $P, Q, R, S \in \mathbb{C}[x_A, x_B]$ and
                \begin{enumerate}[label=(\arabic*)]
                    \item $\text{deg}(P) \preccurlyeq (m, n - m)$ and $\text{deg}(Q) \preccurlyeq (m - 1, n - m - 1)$ and $\text{deg}(R) \preccurlyeq (m - 1, n - m)$ and $\text{deg}(S) \preccurlyeq (m, n - m - 1)$.
                    \item $P$ has negation parity-$(m, n - m) \pmod{2}$ and $Q$ has negation parity-$(m - 1, n - m - 1) \pmod{2}$ and $R$ has negation parity-$(m - 1, n - m) \pmod{2}$ and $S$ has negation parity-$(m, n - m - 1) \pmod{2}$.
                    \item For all $x_A, x_B \in [-1, 1]^2$ the relation
                        \begin{align}
                            |P|^2 + (1 - x_A^2)(1 - x_B^2)|Q|^2 &+ (1 - x_A^2)|R|^2 + (1 - x_B^2)|S|^2
                            \nonumber\\
                            &+ \sqrt{1 - x_A^2}\sqrt{1 - x_B^2}(PQ + P^*Q^* + RS + R^*S^*) = 1,\label{eq:det_relation_x}
                        \end{align}
                    holds, where $P, Q, R, S$ are in $\mathbb{C}[x_A, x_B]$.
                    \item The statement of Conjecture~\ref{conj:m_qsp_factorability} holds as given.
                \end{enumerate}
            
             Proof follows from mapping Theorem~\ref{thm:major_m_qsp_properties} to the $x$ picture. We give this theorem in this form (the $x$-picture) mostly to match with the form of the standard QSP theorems. Unfortunately proving things in this picture is not clean, mostly because the domain is a square in $x$ space due to branch cuts in the square root function, rather than a natural toroidal domain in the Laurent picture.
        \end{lemma}
        
        While Lemma~\ref{lemma:m_qsp_form_x} sits in neat analogy with Theorem~\ref{thm:major_qsp_properties}, it is largely useless for the techniques of positive extensions beyond some opaque versions of Schmüdgen's positivstellensatz \cite{marshall_sos_08}. Consequently from here on we work in the simplified Laurent picture (in other words, choosing to work on the natural torus carved out by the two rotation oracles).
        
        \begin{theorem} \label{thm:major_m_qsp_properties}
            Unitary form of multivariable quantum signal processing (M-QSP). Let $n \in \mathbb{N}$. There exists $\Phi = \{\phi_0, \phi_1, \cdots, \phi_n\} \in \mathbb{R}^{n + 1}$ such that for all $(a, b) \in \mathbb{T}^2$:
                \begin{equation} \label{eq:m_qsp_def}
                    U_{(s, \Phi)}(a, b) = e^{i\phi_0 \sigma_z}\prod_{k = 1}^{n} A^{s_k}(x_A)B^{1 - s_k}(x_B)e^{i\phi_k \sigma_z}
                    = 
                    \begin{pmatrix}
                        P & Q\\
                        -Q^* & P^*
                    \end{pmatrix},
                \end{equation}
            where $A = \mathbb{I}(a + a^{-1})/2 + \sigma_x(a - a^{-1})/2$ and $B = \mathbb{I}(b + b^{-1})/2 + \sigma_x(b - b^{-1})/2$ if and only if $P, Q \in \mathbb{C}[a, b]$ (Laurent polynomials) and
                \begin{enumerate}[label=(\arabic*)]
                    \item $\text{deg}(P) \preccurlyeq (m, n - m)$ and $\text{deg}(Q) \preccurlyeq (m, n - m)$ for $m = |s|$ the Hamming weight of $s$.
                    \item $P$ has parity-$n \pmod{2}$ under $(a, b) \mapsto (a^{-1}, b^{-1})$ and $Q$ has parity-$(n - 1) \pmod{2}$ under $(a, b) \mapsto (a^{-1}, b^{-1})$.
                    \item $P$ has parity $m \pmod{2}$ under $a \mapsto -a$ and parity $m - n \pmod{2}$ under $b \mapsto -b$ and Q has parity $m - 1 \pmod{2}$ under $a \mapsto -a$ and parity $n - m - 1 \pmod{2}$ under $b \mapsto -b$.
                    \item For all $(a, b) \in \mathbb{T}^2$ the relation $|P|^2 + |Q|^2 = 1$ holds.
                    \item The statement of Conjecture~\ref{conj:m_qsp_factorability} holds as given.
                \end{enumerate}
            This result is posed similarly to that of Theorem~\ref{thm:major_qsp_properties}, and its proof is similar up to the use of Conjecture~\ref{conj:m_qsp_factorability}, as shown in Appendix~\ref{appx:m_qsp_proofs}.
        \end{theorem}
        
        \begin{definition} \label{def:stable_polynomials}
            Stable polynomials in one and many variables. A polynomial $p(z_1, z_2, \cdots, z_n)$ is said to be stable if the polynomial does not have zeros in the multi-disk
                \begin{equation}
                    \mathbb{D}^n = \{(z_1, z_2, \cdots, z_n) \in \mathbb{C}^n \text{ s.t. } |z_1| \leq 1, |z_2| \leq 1, \cdots, |z_n| \leq 1\}.
                \end{equation}
            There exist other definitions where one considers polynomials with no zeros in the multi upper half-plane, but these are equivalent up to conformal maps. This family of polynomials is quite restricted, though they are ubiquitous in classical signal processing and control theory.
            
            In the single and two variable case it is further known that all such polynomials are determinantal, i.e., that for every $p(z_1, z_2)$ of degree $(n_1, n_2)$ with no zeros in the bidisk and $p(0, 0) = 1$, one can write $p(z_1, z_2) = \det{(I - KZ)}$, where $Z$ is an $(n_1 + n_2)\times(n_1 + n_2)$ diagonal matrix with $z_1$ or $z_2$ in each diagonal position, and $K$ a contraction. Here a contraction is a linear map with $\lVert K \rVert_{\infty} < 1$ \cite{geronimo_stable_06, grinshpan_stable_16}.
        \end{definition}
        
        \begin{theorem} \label{thm:partial_m_qsp_properties}
            Let $n \in \mathbb{N}$ fixed. Let $\tilde{P}, \tilde{Q} \in \mathbb{C}[a, b]$ (Laurent) such that, for all $(a, b) \in \mathbb{T}^2$, $\tilde{P}(a, b)$ and $\tilde{Q}(a, b) \in \mathbb{R}$. There exist some \emph{stable} $P, Q \in \mathbb{C}[a, b]$ (Laurent) satisfying conditions (1-4) of Theorem~\ref{thm:major_m_qsp_properties} such that $\tilde{P} = \Re(P)$ and $\tilde{Q} = \Re(Q)$ if and only if $\tilde{P}, \tilde{Q}$ satisfy conditions (1-3) of Theorem~\ref{thm:major_m_qsp_properties} and for all $a, b \in \mathbb{T}^2$
                \begin{equation}
                    f(a, b) = 1 - |\tilde{P}|^2 - |\tilde{Q}|^2 > 0,
                \end{equation}
            where $f(a, b)$ is a Laurent polynomial, (note the strict inequality) \emph{and} the additional property holds that the doubly-indexed Toeplitz matrix $c_{u - v}$ (defined in Appendix~\ref{appx:m_qsp_proofs}) populated entirely by differences of the (finitely many) Fourier components of $1/f$ satisfies
                \begin{equation} \label{eq:low_rank_condition}
                    \left[(c_{u - v})_{u, v \in \Lambda\setminus\{0, 0\}}\right]^{-1}_{\subalign{&\{1, 2, \cdots, m\}\times\{0\} \\ &\{0\}\times\{1, 2, \cdots, n - m\}}} = 0
                \end{equation}
            where $\Lambda = \{0, \cdots m\}\times\{0, \cdots n - m\}$ for $m = |s|$ as before. In this case $f(a, b)$ can be written as the square of a \emph{stable} Laurent polynomial. The same result holds if we consider purely imaginary $\tilde{P}, \tilde{Q}$ on $\mathbb{T}^2$, and we can choose either $\tilde{P} \equiv 0$ or $\tilde{Q} \equiv 0$ if desired for simplicity. If the conditions hold, the satisfying $P, Q$ can be computed efficiently in $n$. For an explicit construction of the Toeplitz matrix $c_{u - v}$ refer to the proof of this theorem in Appendix~\ref{appx:m_qsp_proofs}, which relies on intricate results in \cite{mv_frt_04}. Note also that this result is independent of Conjecture~\ref{conj:m_qsp_factorability}.
        \end{theorem}
        
        \begin{conjecture} \label{conj:m_qsp_factorability}
            The FRT = QSP conjecture. In both standard QSP and M-QSP, one is interested in the properties of the single-qubit unitary embedding certain symmetric (trigonometric \& Laurent) polynomial transformations, i.e.,
                \begin{equation} \label{eq:unitary_form_conjecture}
                    U = 
                    \begin{pmatrix}
                        P & Q\\
                        -Q^* & P^*
                    \end{pmatrix},
                \end{equation}
            where $P, Q$ are one- or two-variable polynomials in each setting respectively. This matrix has determinant $1$; this imposes relations between the coefficients of $P$ and $Q$. Specifically, it means that certain sums of products of coefficients constituting the polynomial defining the determinant, $|P|^2 + |Q|^2 = 1$, must be zero. In standard QSP, we make use of the fact that this implies $P_{d_A} = e^{i\phi_A}Q_{d_A}$ where $P_k$ is the coefficient of $a^k$ in $P$ and so on, and where $d_A$ is the largest positive degree of $a$ present. It is this simple constraint that permits QSP phase read-off (the QSP phase in fact relates closely to $\phi_A$).
            
            In the multivariable case the required property no longer holds manifestly, and must be proven; crucially, and this is why we term this the FRT = QSP conjecture, in the single-variable setting the use of the Fejér-Riesz theorem (FRT) in Theorem~\ref{thm:partial_qsp_properties} allows QSP phase read-off precisely because all unitaries of the form given in Eq.~\ref{eq:unitary_form_conjecture} satisfy $P_{d_A} = e^{i\phi_A}Q_{d_A}$, and thus all FRT-generated completions lead to achievable QSP unitaries. In the multivariable setting this same FRT = QSP equivalence corresponds to a strong property of unitary matrices with multivariable polynomials as elements. This property is the conjecture of interest, and there are reasons discussed in Remark~\ref{rem:conjecture_frt_relation} for its reasonableness. At an intuitive level, it is the statement that unitary matrix completions with minimal symmetries and constraints on their elements (basically parity and norm) are always products of low-degree unitary iterates. This is, in again another sense, an extension of the FRT for unitary matrix completions. We depict some of these relationships further in Fig.~\ref{fig:conjecture_flow_diagram}.
            
            Having developed a little motivation, we give a precise statement for FRT = QSP. Given a unitary matrix satisfying the conditions (1-4) of Theorem~\ref{thm:major_m_qsp_properties} (these are the required symmetries and constraints), the single-variable Laurent polynomial coefficients of $P, Q$ satisfy one or both of the following relations for $\phi_A, \phi_B \in \mathbb{R}$:
                \begin{align}
                    \sum_{k = -d_B}^{d_B} P_{d_A, k} b^{k} &= 
                    e^{i\phi_A}\left[\sum_{k = -d_B}^{d_B} Q_{d_A, k} b^{k}\right]\\[0.3em]
                    \sum_{k = -d_A}^{d_A} P_{k, d_B} a^{k} &= 
                    e^{i\phi_B}\left[\sum_{k = -d_A}^{d_A} Q_{k, d_B} a^{k}\right].
                \end{align}
            Here $d_A, d_B$ are the maximal positive degree of $a, b$ respectively appearing in $P, Q$. In other words, either of the two single-variable Laurent polynomial coefficients of maximal degree terms in the other variable must differ only by an overall phase. As shorthand we will denote this set of conditions by the following: 
                \begin{align}
                    P_{d_A}(b) &= e^{i\phi_A}Q_{d_A}(b)\\
                    P_{d_B}(a) &= e^{i\phi_B}Q_{d_B}(a).
                \end{align}
            Here $P_{d_A}(b)$ denotes the single variable Laurent polynomial (in $b$) coefficient of the $a^{d_A}$ term of $P$, and analogously for the other terms.
        \end{conjecture}
        
        Simple inspection of a small subset of the equations relating the two polynomials $P, Q$ induced by condition (4) of Theorem~\ref{thm:major_m_qsp_properties} indicates that these polynomials could in general differ by an overall phase \emph{and a conjugation of some subset of their root multisets}, as opposed to the requirement of Conjecture~\ref{conj:m_qsp_factorability}. Consequently this conjecture is strictly stronger than what is induced by the highest order terms of condition (4), though there are reasons to suspect its truth as discussed in Sec.~\ref{sec:discussion} and Remark~\ref{rem:conjecture_frt_relation}.
        
        \begin{remark} \label{rem:conjecture_frt_relation}
            As discussed above, Conjecture~\ref{conj:m_qsp_factorability} (FRT = QSP) is necessary and sufficient for constructive theorems of M-QSP that rely only on the multivariable Fejér-Riesz theorem. Put colloquially, it is the statement that unitary matrix completions furnished by the FRT always allow themselves to be broken into products of low-degree unitary iterates (in this case the $X, Z$ rotations of QSP). This is a strong condition, but we discuss intuition for why it might hold, as well as the usefulness of the FRT for understanding the M-QSP ansatz (Def.~\ref{def:m_qsp}) even when this conjecture does not hold. We also show that M-QSP suggests useful elaborations on the FRT.
            
            Separate in-progress numerical work shows that the conjecture holds for small-length M-QSP protocols; it is the lack of a simple inductive or otherwise bootstrappable argument that is of concern. That said, the path towards a solution may be obvious when rephrased in terms of other subfields of algebraic geometry; methods towards showing the conjecture involve looking at each of the $\mathcal{O}(d_A d_B)$ homogenous polynomial equations relating the coefficients of $P, Q$ according to condition (4) in Theorem~\ref{thm:major_m_qsp_properties}. The study of simultaneous systems of multivariable polynomial equations is rich, though full of its own conjectures.
            
            Moreover, by simple counting arguments, we see that the coefficients of the leading degree terms of the pairs $P_{d_A}(b), Q_{d_A}(b)$ and $P_{d_B}(a), Q_{d_B}(a)$ in M-QSP are far more constrained than in the single-variable case; namely each of the $\mathcal{O}(d_B)$ or $\mathcal{O}(d_A)$ scalar coefficients respectively of these single-variable polynomials itself enters into $\mathcal{O}(d_A d_B)$ other relations for lower-order homogeneous polynomial equations constituting the determinant. Consequently it is not unreasonable that the roots of at least one of these pairs of polynomials might be forced to differ by no non-trivial conjugations, toward contradiction of the unitarity of the overall transform. In turn, such constructions would provide an entirely new setting for realizing even \emph{non-stable} matrix completions, for which FRT-like theorems have had nothing to say \cite{mv_frt_04}.
        \end{remark}
        
        The FRT = QSP conjecture is not all bad news; note that any \emph{stable} polynomial transform possible to embed in an M-QSP protocol will, by the uniqueness promise \cite{mv_frt_04} of the multivariable Fejér-Riesz theorem up to this stability, lead to a valid M-QSP unitary whose corresponding phases can be efficiently determined by Corollary~\ref{cor:m_qsp_phase_readoff}. Moreover, as is the case in standard QSP, often one wishes not to determine a polynomial transformation directly and then determine its phases, but optimize over the space of possible QSP transformations toward a well-functioning protocol; nothing in the construction of M-QSP prevents this, and results in multivariable approximation theory, as given in Sec.~\ref{sec:discussion}, suggest no major hits to efficiency either.
        
        \begin{corollary} \label{cor:m_qsp_phase_readoff}
            Given a unitary which satisfies conditions (1-4) of Theorem~\ref{thm:major_m_qsp_properties} and is promised to have the form of Definition~\ref{def:m_qsp}, the bitstring $s$ and M-QSP phases $\Phi$ can be efficiently determined by a classical algorithm given $P, Q$. Note that the statement of this corollary trivially circumvents Conjecture~\ref{conj:m_qsp_factorability}. Proof is given in Appendix~\ref{appx:m_qsp_proofs}, within the proof of Theorem~\ref{thm:major_m_qsp_properties}.
        \end{corollary}
    
    This completes a minimal treatment of M-QSP. Evidently its defining theorems are not as succinct as their single-variable analogues. This is for two reasons: (1) the addition of the non-intuitive requirements of the multivariable Fejér-Riesz theorem in Theorem~\ref{thm:partial_m_qsp_properties}, and the reliance on Conjecture~\ref{conj:m_qsp_factorability} in Theorem~\ref{thm:major_m_qsp_properties}. That said, there is positive news for these theorems. Firstly, the initial hurdle of whether the exponentially-many possible orderings of $A, B$ iterates (determined by the bitstring $s$ of Definition~\ref{def:m_qsp}) prevent efficient phase read-off on principle is shown to be a non-issue by Corollary~\ref{cor:m_qsp_phase_readoff}. Secondly, Conjecture~\ref{conj:m_qsp_factorability} is simple to state and has a reasonable chance of being verified or disproven by counterexample through techniques in the theory of systems of multivariable polynomial equations. Finally, the uniqueness guarantee of multivariable Fejér-Riesz theorem shows that if any sufficient condition for the embeddability of a \emph{stable} multivariable polynomial transform in an M-QSP protocol is found, that this theorem can immediately be used to compute completions of unitary matrices, leading to efficient M-QSP phase read-off. Many paths are possible to close the loop on a full characterization of M-QSP.
    
    Nevertheless, the necessary and sufficient conditions given in Eq.~\ref{eq:low_rank_condition} are opaque, and the types of functions with compact Fourier support that satisfy them are hard to visualize. Sec.~\ref{sec:discussion} is in devotion to these caveats. However, while the space of achievable multivariable polynomials of a given degree with M-QSP is not all possible polynomials of said degree up to bound and parity constraints, we also give simple arguments for why this should not even have been expected, and how slightly increasing the degree may resolve these issues. Moreover, the numerical performance of algorithms optimizing over M-QSP protocols approximating interesting multivariable functions should be relatively good up to known results in multivariable approximation theory.

\section{Worked examples} \label{sec:worked_examples}

    That the M-QSP ansatz can generate complicated multivariable transformations of the eigenvalues of two commuting operators may be intuitively clear, but it is worthwhile to provide a couple worked examples and visualizations to this effect. In this section we briefly examine two choices of parameterization for M-QSP protocols in which the embedded polynomials have closed form, and discuss their utility.
        \begin{enumerate}[label=(\arabic*)]
            \item Trivial M-QSP. Recalling that the bit string $s$ indicates the order in which iterates are applied, take $s$ to be the alternating string of length $2n$, i.e., $s = [01]^n$ (where this parenthetical shorthand indicates repetition of the length-two bit-string). For the QSP phases $\Phi$ consider the all-zeros list $\Phi = \{0, 0, \cdots, 0\}$ of length $2n + 1$. The astute observer may note that in this setting the ordering of $s$ does not matter.
            \item XYZ M-QSP. Take $s$ indicating the order in which iterates are applied to again be the alternating string of length $2n$. Take the $k$-th element of $\Phi$ to be $(-1)^{k} \pi/4$, i.e., $\Phi = \{\pi/4, -\pi/4, \cdots, -\pi/4, \pi/4\}$ of length $2n + 1$.
        \end{enumerate}
    We claim that these two ansätze give two useful classes of M-QSP protocols, and that the latter can be used to answer certain promise problems in noiseless settings which would have had no obvious deterministic quantum solution without the mechanisms of M-QSP. In fact, much like in the single-variable setting, these protocols embed polynomial transformations closely related to the Chebyshev polynomials, which are ubiquitous in the theory of efficient functional approximation. While proposals for multivariable generalizations to Chebyshev polynomials are diverse \cite{lidl_72, dunkl_orthogonal_14, beerends_chebyshev_91}, and the map between $\Phi$ and the expansions of $P, Q$ in a Chebyshev basis is highly non-trivial even for standard QSP, it is useful that such simply described countably infinite families of multivariable polynomials can be achieved. It remains for future study to determine if M-QSP suggests an alternative construction for Chebyshev polynomials over many variables. One can observe that the transformations appearing in the top-left ($P$) and top-right ($Q$) corners of unitaries described in Lemmas~\ref{lemma:trivial_qsp} and \ref{lemma:xyz_qsp} are not only \emph{not} polynomials in the variables $x_a = (a + a^{-1})/2$ and $x_b = (b + b^{-1})/2$, but do not satisfy the well-known orthogonality relations that define Chebyshev polynomials; nevertheless, these are still real (Laurent) polynomials of bounded magnitude on the torus, and achieve the same maximal derivative (i.e., proportional to their degree) properties expected of Chebyshev polynomials.
    
        \begin{lemma} \label{lemma:trivial_qsp}
            Trivial M-QSP. For so-called trivial M-QSP protocols defined in the previous paragraph, the polynomials $P, Q$ defining the resulting unitary have the following form for fixed $n$:
                \begin{align}
                    P(a, b) &= 
                    T_{n}\left[\frac{1}{2}\left(a + \frac{1}{a}\right)\right]
                    T_{n}\left[\frac{1}{2}\left(b + \frac{1}{b}\right)\right]\nonumber\\[0.5em]
                    &+ 
                    \frac{1}{4}\left(a - \frac{1}{a}\right)\left(b - \frac{1}{b}\right)
                    U_{n-1}\left[\frac{1}{2}\left(a + \frac{1}{a}\right)\right]
                    U_{n-1}\left[\frac{1}{2}\left(b + \frac{1}{b}\right)\right],\\[0.5em]
                    Q(a, b) &= 
                    \frac{1}{2}
                    \left(b - \frac{1}{b}\right)
                    T_{n}\left[\frac{1}{2}\left(a + \frac{1}{a}\right)\right]
                    U_{n-1}\left[\frac{1}{2}\left(b + \frac{1}{b}\right)\right]\nonumber\\[0.5em]
                    &+
                    \frac{1}{2}
                    \left(a - \frac{1}{a}\right)
                    T_{n}\left[\frac{1}{2}\left(b + \frac{1}{b}\right)\right]
                    U_{n-1}\left[\frac{1}{2}\left(a + \frac{1}{a}\right)\right],
                \end{align}
            where $T_{k}(x)$ and $U_{k}(x)$ are the $k$-th Chebyshev polynomials of the first and second kind evaluated on $x$. It is a small fun exercise to show that these polynomials must be Laurent polynomials in $(ab)$, exhibiting no cross terms of $a, b$ with differing exponents. Note that $|P|^2 + |Q|^2 = 1$ must hold, and thus each of these polynomials is bounded in magnitude by $1$ on $\mathbb{T}^2$.
        \end{lemma}
        
        \begin{lemma} \label{lemma:xyz_qsp}
            XYZ M-QSP. For the XYZ M-QSP protocol defined in the previous paragraph, the polynomials $P, Q$ defining the resulting unitary have the following form for fixed $n$:
                \begin{align}
                    P(a, b) &= T_{n}\left[\frac{1}{4}\left(a + \frac{1}{a}\right)\left(b + \frac{1}{b}\right)\right] + \frac{i}{4}\left(a - \frac{1}{a}\right)\left(b - \frac{1}{b}\right) U_{n - 1}\left[\frac{1}{4}\left(a + \frac{1}{a}\right)\left(b + \frac{1}{b}\right)\right],\\[0.5em]
                    Q(a, b) &= \frac{1}{4}
                    \left[
                    \left(a - \frac{1}{a}\right)\left(b + \frac{1}{b}\right)
                    -
                    i\left(a + \frac{1}{a}\right)\left(b - \frac{1}{b}\right)
                    \right]
                    U_{n-1}\left[\frac{1}{4}\left(a + \frac{1}{a}\right)\left(b + \frac{1}{b}\right)\right],
                \end{align}
            where $T_{k}(x)$ and $U_{k}(x)$ are the $k$-th Chebyshev polynomials of the first and second kind evaluated on $x$. Note that in most instances these polynomials are simpler in the $\theta$ picture, as $(a \pm a^{-1})$ and $(b \pm b^{-1})$ have nicer expressions as cosine and sine of $\theta_a, \theta_b$ for $a = e^{i\theta_a}, b = e^{i\theta_b}$. The proof of this fact is by simple inspection according to the known recurrence relation for the Chebyshev polynomials. Note that $|P|^2 + |Q|^2 = 1$ must hold, and thus each of these polynomials is bounded in magnitude by $1$ on $\mathbb{T}^2$.
            
            Note that we call this XYZ M-QSP because the chosen phases can be seen to conjugate one subset of the iterates to change them from $X$ to $Y$ rotations on the Bloch sphere,
                \begin{equation}
                    e^{i\theta_a \sigma_x} \mapsto e^{-i\sigma_z\pi/4}e^{i\theta_a \sigma_x}e^{i\sigma_z\pi/4} = e^{i\theta_a \sigma_y},
                \end{equation}
            and thus the ansatz looks like alternating $X$ and $Y$ rotations by $\theta_a, \theta_b$.
        \end{lemma}
        
\begin{figure}[htpb]
            \centering
            \includegraphics[width=1.0\textwidth]{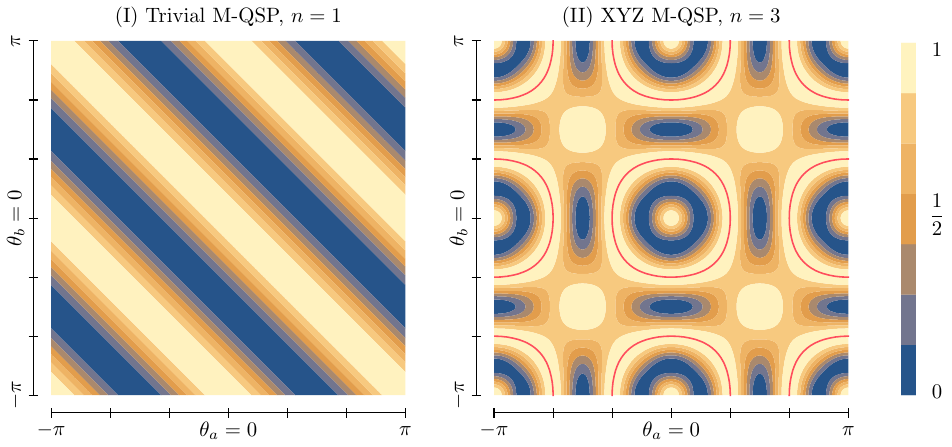}
            \caption{Contour plots of $|P(a, b)|^2$ for the (I) trivial M-QSP protocol with $n = 1$ and the (II) XYZ M-QSP protocol with $n = 3$. Note that by definition that $P$ is bounded in magnitude by $1$. Here $\theta_a, \theta_b$ satisfy $\cos{\theta_a} = (a + a^{-1})/2$ and $\cos{\theta_b} = (b + b^{-1})/2$ and analogously for sine. The toroidal nature of the domain is evident: the plots are also symmetric under $\theta_a \mapsto -\theta_a$, equivalently inversion parity $a \mapsto a^{-1}$, as well as $\theta_a \mapsto \theta_a + \pi$, equivalently negation parity $a \mapsto -a$, and likewise for the $b$ variables. The red line overlay in (II) represents the relation $4\cos^2{\theta_a}\cos^2{\theta_b} = 1$, as discussed in the text. Note finally that (I) is a function purely of $(\theta_a + \theta_b)$ as expected.}
            \label{fig:worked_examples}
        \end{figure}

        We can make a few cursory observations about the polynomial transformations embedded by these ansätze. For trivial M-QSP, the ordering of iterates cannot matter, and thus any embedded transformation must also be a polynomial purely of $(ab)$ as a single-variable; this is depicted in Fig.~\ref{fig:worked_examples}. Likewise, any time a product of $A, B$ iterates is repeated without intervening $Z$ rotations, a corresponding reduction to a single-variable standard QSP protocol is possible. In contrast, the polynomial transforms of XYZ M-QSP for $n \geq 2$ do not factor in this neat way, and we can come up with somewhat contrived but interesting promise problems for which XYZ M-QSP provides a neat solution.
        
        \begin{problem} \label{prob:multi_channel_discrimination}
            Consider the following multi-channel discrimination problem. A querent is given free access to two oracles $A = e^{i\theta_a}$ and $B = e^{i\theta_b}$ and is told that one of the two following scenarios holds:
                \begin{enumerate}[label=(\arabic*)]
                    \item $\theta_a, \theta_b$ are from the four-element set $\{\{0, \pm \pi/2\}\}\cup\{\{\pm\pi/2, 0\}\}$.
                    \item $\theta_a, \theta_b$ satisfy the relation $4\cos^2{\theta_a}\cos^{2}{\theta_b} = 1$.
                \end{enumerate}
            It can be shown that these two cases are disjoint. Moreover, this discrimination problem can be solved deterministically in $6$ total queries using M-QSP (shown in (II) of Fig.~\ref{fig:worked_examples}), while for two quantum parties given access to $A, B$ separately, there is no such algorithm, for the same reasons as those discussed in \cite{noisy_channel_rossi_22, rossi_qht_21}. Additionally, while not discussed here, this query complexity persists even in the presence of small noise \cite{noisy_channel_rossi_22}. That is, no two quantum computing parties connected only by classical communication and sharing no entanglement, given access to one each among the oracles in this problem, can decide this problem deterministically with zero error, nor can this problem be reduced to single-variable QSP by substitution. Consequently M-QSP has permitted the deterministic computation of a structured \emph{joint} property that is not related to sums or differences of encoded signals (i.e., by reduction to single-variable QSP in $a^n b^m$ for integers $n, m$). Here, M-QSP both outperforms serial application of QSP protocols on two oracles individually, and does so for a highly non-trivial functional relation.
            
            Intuitively, note that this advantage relies crucially on an inability to decompose the discrimination regions depicted in Figure~\ref{fig:worked_examples} into blocks on which the circuit output should be constant for a given discrimination problem. Consequently two parties, each given access to one among the two oracles, cannot with certainty disambiguate these joint properties by projecting onto either the $\theta_a$ or $\theta_b$ axis, and classically computing the function outside the quantum circuit.
        \end{problem}
        
        \begin{remark}
            Note that for trivial M-QSP protocols the iterates $A, B$ commute; consequently the embedded functions are single-variable Laurent polynomials in the joint variable $ab$. Clearly such polynomials are not stable, and thus these completions are not furnished by Theorem~\ref{thm:m_frt}. However, through this single-variable reduction these completions \emph{are enabled} by the simpler Fejér-Riesz lemma. It remains an open question whether the application of Theorem~\ref{thm:m_frt} fails only in the settings where a single-variable reduction similar to this case exists. If this is the case, one could consider some non-trivial class M-QSP\textsuperscript{$\star$} of multivariable protocols which admit no single-variable reduction, and thus always (up to small perturbation) permit the application of the FRT as given. This would represent a substantial elaboration on the theory of unitary matrix completions.
        \end{remark}
        
        The central message from this section is a precise point about the utility of M-QSP for computing joint functions of oracular parameters. While such functions could always in principle be computed approximately classically \emph{outside} a quantum circuit by first estimating the relevant parameters tomographically, this can be extremely costly, and may provide more information than a querent wishes to know. As with standard QSP, M-QSP allows precise and \emph{coherent} control over both subsystem dynamics and information extracted by measurement; known structure of the data encoded in the oracle can be leveraged into query complexity savings, written in the language of functional approximation theory. If an M-QSP computation is oblivious to individual, expensive-to-determine properties of each oracle, then it can avoid lower bounds for computing these unimportant properties! Its functional transforms can be carefully controlled, defined by interpolatory or approximative constraints, and cascaded in the application of many QSP protocols without unnecessary measurement and amplification.

\section{Discussion} \label{sec:discussion}
    
    In this work we have developed a theory of multivariable quantum signal processing (M-QSP) in two variables, and shown a variety of results about its properties through direct analogies to theorems of standard QSP. Specifically, we give Theorem~\ref{thm:major_m_qsp_properties}, which shows that up to Conjecture~\ref{conj:m_qsp_factorability}, M-QSP polynomial transforms are only as constrained as those of standard QSP. Moreover, we show in Corollary~\ref{cor:m_qsp_phase_readoff} that given a valid M-QSP protocol there is no obstacle to determining its constitutive real phases $\Phi$ and the order $s$ in which these oracles are applied. Finally, we show Theorem~\ref{thm:partial_m_qsp_properties}: that given a partially-defined M-QSP protocol, the existence of a \emph{stable} completion of said protocol relies solely on the guarantees of a multivariable Fejér-Riesz theorem. This result connects questions in QSP-like algorithms to nullstellensätze, furnishes alternative proofs for results in standard QSP, and opens a variety of concrete questions involving matrix completions in a quantum information setting.
    
    The remaining purpose of this section is fourfold: (1) state caveats related to M-QSP insofar as its properties and guarantees have not been fully characterized, (2) discuss the numerical outlook for M-QSP, (3) give a brief overview of how M-QSP informs a lifted, many-qubit M-QSVT, and finally (4) state avenues of ongoing research and their basal open questions.
    
    \subsection{Caveats and reminders}
        
        The story of M-QSP is not completely resolved in this work. The results in previous sections provide only a partial characterization of the expressive powers of the M-QSP ansatz, and do so in some cases with respect to specific conjectures and caveats. Here we give succinct reminders tethered to these limitations.
            \begin{enumerate}[label=(\arabic*)]
                \item The space of embeddable polynomial transforms in M-QSP is smaller than the space of polynomials with bounded norm and definite parity up to a given degree (as opposed to standard QSP).
                
                We argue that one should not have expected to be able to achieve all polynomial transformations in M-QSP as permitted by the norm and parity constraints given in Theorem~\ref{thm:partial_qsp_properties}. This a simple counting argument: for a multivariable polynomial of degree $(n, m)$ with definite parity and bounded norm on the bitorus, there exist $\mathcal{O}(nm)$ possible small independent perturbations of the coefficients which preserve these properties, but only $\mathcal{O}(n + m)$ real QSP phases parameterizing M-QSP ansätze. It is precisely that the achievable transforms are some low rank subspace of this space of norm-bounded definite-parity transforms that the multivariable Fejér-Riesz theorem's statement captures.
                
                \item The multivariable Fejér-Riesz theorem (FRT) can determine if there exist unitary completions for partially defined polynomial transformations if and only if those completions are \emph{stable} polynomial transforms.
                
                Stable polynomials in one and two variables, as discussed in Definition~\ref{def:stable_polynomials}, have special properties. It is also known that there exist polynomial transforms achieveable by M-QSP which are \emph{not stable}, and thus for which the FRT can say little in the two-variable case. It is an interesting question whether such instability implies the ability to reduce the corresponding transformation to a single-variable setting (or else be infinitesimally perturbed to a stable setting). In the single-variable case, such unstable polynomial embeddings can always be converted to stable ones as all roots are isolated; a similar transformation for the multivariable case is not possible in general \cite{geronimo_stable_06}.
                
                The conjecture (FRT = QSP in Conjecture~\ref{conj:m_qsp_factorability}) that, like in the single-variable case, the multivariable FRT is the only non-trivial mechanism underlying M-QSP, has a succinct statement. Nevertheless, the methods to prove this statement depend on the existence of solutions to simultaneous multivariable polynomial equations, which is a hard problem in general. Any theory of M-QSP must overcome this difficulty.
                
                \item The guarantees of M-QSP critically rely on the two queryable oracles having the same eigenbasis. Consequently lifting M-QSP to multiple qubits requires identical singular vectors for the relevant block encoded operators.
                
                While this restriction is unfortunate, it is not unexpected; considering a theorem of QSP or QSVT for general interleaved non-commuting operators not only exposes us to pathological cases where the oracle set is complete for unitary approximation or quantum computation, but destroys the assumptions of Jordan's lemma \cite{jordan_75} (that two interleaved rotations or reflections select invariant subspaces); such results fundamentally enable QSVT \cite{gslw_19}, and cannot be abandoned without drastically altering the simple utility of the ansatz.
            \end{enumerate}
            
    \subsection{The numerical outlook for M-QSP}
    
        While the theoretical tools from algebraic geometry to understand multivariable polynomials are necessarily not as strong as their single variable analogues \cite{mv_frt_04}, the theory of multivariable polynomial approximation and interpolation is well-developed and in general spells good news for those looking to approach M-QSP from a numerical perspective. We give known results in multivariable approximation theory, showing that there is no fundamental barrier to multivariable analogues for the multiple \emph{classical} subroutines that are used in standard QSP.
    
            \begin{enumerate}[label=(\arabic*)]
                \item Multivariable Stone-Weierstrass theorems. Most basically, it is known that multivariable trigonometric polynomials are dense in the space of continuous functions on the multitorus, just as in the single-variable setting \cite{polya_szego_analysis}.
                \item Multivariable Jackson-type theorems. Jackson's theorems or Jackson's inequalities relate the smoothness of a function and the required degree of a trigonometric polynomial approximation to a desired uniform error. Such theorems exist in the multivariable case \cite{newman_jackson_64, ragozin_jackson_70} and have fundamentally the same character as in the single-variable case, meaning the required degree's best-case polylogarithmic dependence on approximative error achieved by QSP is not forbidden in the M-QSP setting. That said, directly comparing degree in the single-variable and multivariable context has its own caveats \cite{trefethen_hypercube_17}.
                \item Multivariable Remez-type or Parks-McClellan algorithms. Beyond ensuring that there exist good trigonometric polynomial approximations to desired continuous functions, much work related to QSP has centered on good classical algorithms for efficiently computing said polynomial approximations, derived from the well-known signal processing Remez-type/Parks-McClellan algorithms \cite{remez, grenez, parks_mcclellan_72}. These algorithms have multivariable counterparts with similar performance as in the single-variable case \cite{watson_m_remez_75}, although their theoretical guarantees are less well-understood.
                \item Incorporating Fejér-Riesz constraints. If one seeks the stable factorizations guaranteed in \cite{mv_frt_04}, one may worry that optimizing over polynomials which satsify the constraint given in Theorem~\ref{thm:partial_m_qsp_properties} may itself be difficult. That said, numerical work in \cite{mv_frt_04, woerdeman_numerical_03, hachez_approx_05} each support that this constraint leads to well-defined semi-definite programming problems that, while not fully characterized, appear empirically compatible with common classical optimization algorithms discussed above.
                \item Optimizing over QSP protocols. Recent research in standard QSP has shown that imposing symmetries on QSP phases leads to improved performance for numerical optimization over QSP protocols, as well as guarantees of convexity of the search space under well-defined constraints on the desired embedded functional transform \cite{sym_qsp_21}. While we have observed evidence for the benefits of similar symmetrization for M-QSP in our own numerical simulations, formally showing similar guarantees as in the single-variable setting generally is a prime direction for future work.
            \end{enumerate}
        
    \subsection{Lifting M-QSP to M-QSVT}
    
        Much of the interest in QSP-like algorithms stems from their use at the core of algorithms for manipulating the eigenvalues or singular values of larger linear systems embedded in unitary matrices \cite{gslw_19, lloyd_hamiltonian_qsvt_21, lc_19_qubitization}. QSP can be thought of as the special case in which this linear operator is just the single scalar value in the top left of a representation of an SU(2) operator. We briefly review how to lift from QSP to QSVT, and show that M-QSP immediately enables a lifted M-QSVT for pairs of operators that share the same singular vectors (or equivalently, in the case that these operators are square, that they commute).
        
        The purpose of this section is not to exhaustively build a theory of M-QSVT and its uses, but to advertise the powerful fact that anything it is possible to prove about M-QSP leads directly and simply to a lifted, many-qubit context.
        
        As the authors of \cite{gslw_19} succinctly note, interleaving any unitary $U$ with simple phase operators can induce polynomial transforms of the singular values of certain sub-blocks of $U$; the business of QSVT is to explicitly identify these sub-blocks. The preservation of these blocks under repeated interleaving is a corollary of an old result, Jordan's lemma \cite{jordan_75}, and is the reason that any constructive non-commuting version of M-QSP is destined for fundamental problems. Applications of this lemma are ubiquitous in other areas of quantum information, from quantum walks \cite{szegedy_04} to quantum interactive proofs \cite{marriott_watrous_05, fast_qma_jordan_09}, and are useful to understand.
        
        Following the conventions of \cite{gslw_19}, take $\mathcal{H}_U$ some finite-dimensional Hilbert space on which $U$ acts, and $\tilde{\Pi}, \Pi$ orthogonal projectors which locate the linear operator $A$ according to
            \begin{equation}
                A = \tilde{\Pi}U\Pi.
            \end{equation}
        We say here that $U$ block encodes $A$, and this idea is formalized through Sec.~4 of \cite{gslw_19}. Taking $d = \text{rank}{(\Pi)}$, $\tilde{d} = \text{rank}{(\tilde{\Pi})}$, and $d_{\rm min} = \min{(d, \tilde{d})}$, the singular value decomposition of $A$ is simple. Take $\{|\psi_j\rangle\}, j \in [d]$ and $\{|\tilde{\psi}_j\rangle\}, j \in [\tilde{d}]$ to be orthonormal bases for $\text{img}{(\Pi)}$ and $\text{img}{(\tilde{\Pi})}$ respectively, then
            \begin{equation}
                A = \sum_{j = 1}^{d_{\rm min}} \xi_{j}|\tilde{\psi}_j\rangle\langle \psi_j |,
            \end{equation}
        where each of the $\xi_{j}$ are in $[0, 1]$, and $\xi_{j} \geq \xi_{k}$ for $1 \leq j \leq k \leq d_{\rm min}$. The careful work of \cite{gslw_19} shows that the action of $U$ with respect to this basis acts as
            \begin{equation} \label{eq:qsvt_u_decomp}
				U = \cdots \oplus\; \bigoplus_{\xi_{j} \neq 0, 1}
				\begin{bmatrix}
					\xi_j & \sqrt{1 - \xi_j^2}\;\\
					\sqrt{1 - \xi_j^2} & -\xi_j
				\end{bmatrix}^{\mathcal{H}_j}_{\tilde{\mathcal{H}}_j}
				\oplus \cdots,
			\end{equation}
		where the block's superscript $\mathcal{H}_{j}$ and subscript $\tilde{\mathcal{H}}_j$ indicate that it maps from the space spanned by the $|\psi_j\rangle$ to that spanned by the $|\tilde{\psi}_j\rangle$, each of which is a subspace of the original $\mathcal{H}_U$. The additional components in the direct sum, i.e., the action of $U$ on the rest of $\mathcal{H}_U$, can be written explicitly, and correspond to actions of $U$ outside the relevant images of $\Pi, \tilde{\Pi}$; this careful bookkeeping is documented neatly in \cite{gslw_19}. Their construction crucially introduces two further phase operators which can be shown to be easily constructable, namely
		    \begin{align}
		        e^{i\phi(2\Pi - I)} &= \cdots \oplus\; \bigoplus_{\xi_{j} \neq 0, 1}
				\begin{bmatrix}
					e^{i\phi} & 0\;\\
					0 & e^{-i\phi}
				\end{bmatrix}^{\mathcal{H}_j}_{\mathcal{H}_j}
				\oplus \cdots,\label{eq:phase_decomp_1}\\
				e^{i\phi(2\tilde{\Pi} - I)} &= \cdots \oplus\; \bigoplus_{\xi_{j} \neq 0, 1}
				\begin{bmatrix}
					e^{i\phi} & 0\;\\
					0 & e^{-i\phi}
				\end{bmatrix}^{\tilde{\mathcal{H}}_j}_{\tilde{\mathcal{H}}_j}
				\oplus \cdots,\label{eq:phase_decomp_2}
		    \end{align}
		which together with the action of $U$ allow us to recognize interleaved products of these operators as performing effectively (up to substitutions of rotations for reflections) QSP in each of the singular vector subspaces defined by these projectors. In effect, the blocks as shown in the above equations multiply only with each other all thanks to Jordan's lemma, and each of these sub-blocks looks just like a product of single-qubit rotations (or antiunitary reflections, in this setting). In more specific words, drawing from Definition 15 and Theorem 17 of \cite{gslw_19}, the alternating protocol (assuming for the moment the protocol's length $n$ is even)
            \begin{equation}
                U_{\phi} = \prod_{j = 1}^{n/2} \left[e^{i\phi_{2j - 1}(2\Pi - I)}U^{\dag} e^{i\phi_{2j}(2\tilde{\Pi} - I)}U\right],
            \end{equation}
        can be shown to induce a desired polynomial transform in precisely the way QSP does, explicitly that
            \begin{equation}
                \tilde{\Pi}U_{\Phi}\Pi = \sum_{j = 1}^{d_{\rm min}} P(\xi_j)|\tilde{\psi}_j\rangle\langle \psi_j |,
            \end{equation}
        where this polynomial transformation is effectively (again up to a simple map between reflections and rotations) the same one as generated by the standard QSP protocol with QSP phases $\Phi$. All that is required for polynomial transformation of degree $n$ are $n$ uses of $U, U^\dag$, the $\Pi, \tilde{\Pi}$-controlled NOT gates needed for the phase operators, single-qubit phase gates, and a constant number of auxiliary qubits.
        
        The usefulness of this basic construction in lifting M-QSP should be clear; given two linear operators $A_1, A_2$ which have \emph{the same singular vectors} in their singular value decompositions and which are located within unitaries $U_1, U_2$ of the same size, the decompositions given in Eqs.~\ref{eq:qsvt_u_decomp}, \ref{eq:phase_decomp_1}, and \ref{eq:phase_decomp_2} still hold. I.e.,
            \begin{equation}
                A_1 = \tilde{\Pi} U_1 \Pi, \quad A_2 = \tilde{\Pi} U_2 \Pi,
            \end{equation}
        where only if $A_1, A_2$ share singular vectors do the subspaces discussed above remain invariant under operator interleaving. A simpler instance of this phenomenon is when $U_1, U_2$ block encode commuting operators, in which case this condition is obviously satisfied.
        
        Note that this is clearly a restrictive condition, though it should come as no surprise: no analogue of Jordan's lemma exists for interleaved products of more than two operators, as there exist cases in which no non-trivial subspaces would be preserved by such action. Here we choose to preserve the two-dimensional subspaces spanned by the singular vectors, but allow the singular values to differ. Evidently M-QSVT is no harder to construct than M-QSP, and it is left to the interested reader to port results in the latter to the context of the former; any instance of block-encodable commuting operators can be discussed in the terminology of M-QSP!

    \subsection{Ongoing work and open questions}
    
        Finally we outline major open avenues for M-QSP, aimed at theoretical physicists, theoretical computer scientists, those in industry, and pure mathematicians looking to switch subfields. M-QSP, like its single-variable analogue, is ripe for simultaneous analytical and numerical investigation; its utility is centered in the relative simplicity of its defining ansatz in conjunction with the rigorously characterized expressiveness of its generated transformations. This, together with its low resource-overhead, makes it a good candidate for continuing the legacy of QSVT in unifying the current pantheon of quantum algorithms, as well as realizing them on near-term devices.
    
            \begin{enumerate}[label=(\arabic*)]
                \item M-QSP opens the door to considering other alternating ansätze: variations include restrictions (e.g., strictly alternating protocols, protocols with symmetrized phases), as well as elaborations (e.g., arbitrarily many oracles). It is the intent of Sec.~\ref{sec:alg_geo_review} to introduce readers to the mathematical subfields which may in turn inform new, far larger classes and families of QSP- and QSVT-like algorithms. Some of these may not only make use of scalar factorization results, but the diverse families of operator-valued results in the theory of positive extensions \cite{mv_frt_04, op_val_frt_09}.
                \item M-QSP can be used even in the absence of a complete characterization of all circuits stemming from all reasonable ansätze. Indeed, the existence of even a single countably infinite family of embedable polynomial transformations can yields proofs of quantum advantage. Thus empirical research into such families, even by those who have no wish to understand deep takes in algebraic geometry, is worthwhile.
                \item As stated, lifting M-QSP to M-QSVT gives a theory only of commuting block encoded operators; this is necessary in order to preserve the use of Jordan's lemma. Relaxing this constraint in general may be impossible, but investigating situations in which more complicated subspaces are preserved by the interleaving operators used in QSVT may be possible. Novel pure mathematical investigations into variants of Jordan's lemma have great utility for quantum information beyond QSP/QSVT \cite{marriott_watrous_05, fast_qma_jordan_09}, and are a great starting point for foundational work. In turn, these result better inform our understanding of control of subsystem dynamics (a fundamental question in quantum information) and a variety of other periodic circuit ansätze.
                \item Improved Fejér-Riesz theorems for non-stable factorizations. Known results in positive extensions consider stable factorizations, which in the single-variable case is no problem because such polynomials have discrete zeros. While firmly in the realm of pure mathematics, relaxing such theorems to consider both non-stable factorizations and nonnegative (rather than positive) multivariable trigonometric polynomials would greatly impact the theory of quantum algorithms. Indeed, showing the FRT = QSP Conjecture (Conjecture~\ref{conj:m_qsp_factorability}) would provide useful examples of such extensions, and thus add to a critical series of results in algebraic geometry from a pragmatic computational perspective.
            \end{enumerate}
        
        As a final takeaway: QSP-like algorithms derive their utility from their complete characterization of control over subsystem dynamics of unitary evolutions. This is a \emph{bottom-up} approach that avoids the difficulties of circuit ansätze like VQEs or QNNs, whose properties are often difficult to treat rigorously, and which must primarily be investigated numerically or heuristically. M-QSP remains in the vein of QSP's original successful approach—it seeks to take advantage of the usual Feynmanian adage: \emph{there's plenty of room at the bottom}.
            
    \section{Acknowledgements}
    
        The authors would like to thank Sho Sugiura, John Martyn, Patrick Rall, Alexander Zlokapa, and Jordan Docter for helpful discussions.
        ZMR was supported in part by the NSF EPiQC program, and IC was supported in part by the U.S. Department of Energy, Office of Science, National Quantum Information Science Research Centers, and Co-design Center for Quantum Advantage (C2QA) under contract number DE-SC0012704. We also acknowledge NTT Research for their financial and technical support.

\bibliography{main}

\appendix

\section{Extensions to the Fejér-Riesz lemma} \label{appx:m_qsp_proofs}
    
    The purpose of this appendix is threefold: (1) incorporate the single-variable Fejér-Riesz lemma into proofs of main theorems of single-variable QSP, (2) use modified and considerably more involved multivariable versions of this lemma in the proofs of major properties of M-QSP, and (3) generally acquaint the study of QSP-like ansätze with a relevant and well-understood subfield of algebraic geometry. This work is inspired by results which descend from the study of Hilbert's 17\textsuperscript{th} problem. Related statements are, with a little maneuvering, ubiquitous in quantum information, and may offer more insight to the interested researcher than the utilitarian implementation here.

    \subsection{Single-variable setting}
    
        We fulfill a promise to re-prove Theorem~\ref{thm:partial_qsp_properties} in the Laurent picture, and show that it relies entirely on a relatively simple and clean result, the aforementioned Fejér-Riesz lemma; this lemma provides a concrete description of an entire sub-class of positive polynomials. Almost all of the major results in the previous work on QSP outside of its classical subroutines can be reformulated to center on this lemma, and are arguably made cleaner and more compact by this reduction due to the removal of awkward branch cuts in the $x$ picture.
    
        \subsubsection{Related definitions and lemmas}
    
            \begin{lemma} \label{lemma:frt}
                Single-variable Fejér-Riesz lemma (an old result, recalled in \cite{mv_frt_04}). A \emph{single-variable} trigonometric polynomial
                    \begin{equation}
                        f(z) = \sum_{k = -n}^{n} f_{k}z^{k},
                    \end{equation}
                taking non-negative values on $\mathbb{T}$ can always be expressed as the modulus squared of a polynomial of the same degree, i.e., there exists $g(z) = g_{0} + \cdots + g_{n}z^n \in \mathbb{C}[z]$ such that
                    \begin{equation}
                        f(z) = |g(z)|^2 = g(z)g^*(z^{-1}),
                    \end{equation}
                where the degree of $g$ is the same as that of $f$. In fact one can choose $g(z)$ to be \emph{outer}, i.e., $g(z) \neq 0$ for $|z| < 1$, and in the non-singular case, when $f(z) > 0$ for $|z| = 1$, one can choose $g(z)$ to be stable, namely $g(z) \neq 0$ for $|z| \leq 1$. Up to this choice and an overall phase the factorization is unique. A standard proof (among many) of this lemma is found with Theorem 1.1 of \cite{dumitrescu_monograph_07}, and relies only on the fundamental theorem of algebra.
            \end{lemma}
        
        \subsubsection{Alternative proof of Theorem \ref{thm:partial_qsp_properties}}
        
            We modify the statement of Theorem~\ref{thm:partial_qsp_properties} in a way that is amenable to application of the single-variable Fejér-Riesz lemma. This allows the proof of the multivariable analogue of this theorem in Appendix~\ref{appx:m_qsp_proofs} to be more familiar.
            
            Consider, as was given in Theorem~\ref{thm:partial_qsp_properties}, that the real polynomials $\tilde{P}$ and $\tilde{Q}$ satisfy the inequality $|\tilde{P}|^2 + (1 - x^2)|\tilde{Q}|^2 \leq 1$. Consider instead a renaming of the polynomial $\tilde{P}(x)$ by its Laurent picture version $P(z) \in \mathbb{C}[z]$ and the non-polynomial $\sqrt{1 - x^2}\tilde{Q}(x)$ by its Laurent picture version (now truly a Laurent polynomial) $Q(z) = -(i/2)(z - z^{-1})\tilde{Q}(z)$. In other words we consider the matrix
                \begin{equation} \label{eq:laurent_condition}
                    \begin{pmatrix}
                        P(z) + iR(z) & Q(z) + iS(z) \\
                        -Q(z) + iS(z) & P(z) - iR(z)
                    \end{pmatrix},
                \end{equation}
            where $P(z)$ and $Q(z)$ are known and real on $\mathbb{T}^2$, and where we want to determine if there exist $R(z)$ and $S(z)$ (again real on $\mathbb{T}^2$) satisfying conditions (1-2) of the theorem statement as well as the determinant condition
                \begin{equation}
                    P(z)^2 + Q(z)^2 + R(z)^2 + S(z)^2 = 1.
                \end{equation}
            Note that we have used the known even $z \mapsto -z$ parity of each Laurent polynomial to fill out Eq.~\ref{eq:laurent_condition}, and that while each of these polynomials is either real or imaginary on $\mathbb{T}^2$, this does not imply the coefficients of the Laurent polynomial are either all real or all imaginary, only Hermitian as stated previously.
            
            Now define the nonnegative degree-$2n$ Laurent polynomial $U(z) = 1 - P(z)^2 - Q(z)^2$ and apply the Fejér-Riesz lemma (Lemma \ref{lemma:frt}) to yield a stable \emph{real-coefficient} (not real on $\mathbb{T}^2$) polynomial of degree $2n$ of the form
                \begin{equation}
                    T(z) = \sum_{k = 0}^{2n} t_k z^k, \; t_k \in \mathbb{R},
                \end{equation}
            which must satisfy
                \begin{align}
                    U(z) = R(z)^2 + S(z)^2 &= (R(z) + iS(z))(R(z) - iS(z))\nonumber\\ 
                    &= (T(z) z^{-n}) (z^{n} T^*(z^{-1}))\nonumber\\
                    &= (T^\prime(z)) (T^{\prime*}(z^{-1})),
                \end{align}
            where we have added dummy powers of $z$ to make the $T(z)$ guaranteed by the Lemma \ref{lemma:frt} into a degree-$n$ Laurent polynomial $T^\prime(z)$ with real coefficients. Consequently the symmetric and antisymmetric components of this polynomial, with respect to $z \mapsto 1/z$ can be matched with $R(z)$ and $S(z)$ respectively and unambigously (the latter absorbing the factor of $i$). This preserves the desired $z \mapsto -z$ parity of each element, condition (2), the desired degree constraint, condition (1), and finally the determinant constraint, condition (3), which is what we desired to show. Returning to the $x$ picture and pulling out the necessary factor of $-(i/2)(z - z^{-1})$ from $Q(z)$ (possible to pull out because of the fundamental theorem of algebra guaranteeing roots at $z = \pm 1$), we recover the usual $\sqrt{1 - x^2}$ term. Thus, without much extra work, the Fejér-Riesz lemma comes across as the only non-trivial mechanism underlying the reverse problem ($\tilde{P}, \tilde{Q} \mapsto \Phi$) of QSP.
    
    \subsection{Multivariable setting}
        
        We provide proofs of the constitutive theorems of M-QSP. Beyond reference to some major (and complexly derived) theorems in functional analysis, this subsection is self-contained, and aimed toward a simplified analytic presentation. Where indicated we include explicit reference to conjectures and related results as depicted in Fig.~\ref{fig:conjecture_flow_diagram}.
    
        \subsubsection{Related definitions and lemmas}
        
            In the following proofs we consider matrices whose rows and columns are indexed by subsets of $\mathbb{Z}^2$ as described in \cite{mv_frt_04}. For instance, if $U = \{(0, 0), (0, 1), (1, 0)\}$ and $V = \{(2, 1), (2, 2), (2, 3)\}$ then we denote by $C = (c_{u - v})_{u \in U, v \in V}$ the $U \times V$ (i.e., $3\times 3$) matrix
                \begin{equation}
                    C = 
                    \begin{pmatrix}
                        c_{-2,-1} & c_{-2,-2} & c_{-2,-3}\\
                        c_{-1,-1} & c_{-1,-2} & c_{-1,-3}\\
                        c_{-2,0} & c_{-2,-1} & c_{-2,-2}
                    \end{pmatrix},
                \end{equation}
            which evidently indexes elements in the $i, j$ (row and column) position by taking the difference of the $i$-th element of $U$ and the $j$-th element of $V$.
            
            We will usually consider the set $\Lambda = \{0, 1, \cdots, n\}\times\{0, 1, \cdots, m\}$ and its use in generating a $(n + 1)\times(n + 1)$ block Toeplitz matrix whose blocks are themselves $(m + 1)\times(m + 1)$ Toeplitz matrices. This Toeplitz matrix is the result of the multi-indexing procedure discussed previously, e.g., given $c_{0}, c_{1}, \cdots, c_{n}$ one can define $C = c_{i-j}, i,j \in \{0, 1, \cdots, n\}^2$ which has the form
                \begin{equation}
                    C = 
                    \begin{pmatrix}
                        c_{0} & c_{1}^* & \cdots & c_{n}^*\\
                        c_{1} & c_{0} & \ddots & \vdots\\
                        \vdots & \ddots & \ddots & c_{1}^*\\
                        c_{n} & \cdots & c_{1} & c_{0}
                    \end{pmatrix},
                \end{equation}
            where the $c_{k}$ we consider will have the additional Hermitian property $c_{-k} = c_{k}^*$.
            
            \begin{definition}
                Doubly-indexed Toeplitz matrix of Fourier components. Suppose a function $f(a, b): \mathbb{T}^2 \rightarrow \mathbb{C}$ has non-zero Fourier components $\hat{f}(k, l) = c_{kl}$ for $(k, l) \in \{0,1, \cdots, n\}\times\{0, 1, \cdots, m\}$. Then the doubly-indexed Toeplitz matrix $\Gamma$ corresponding to these Fourier components has form
                    \begin{equation} \label{eq:gamma_toeplitz}
                        \Gamma = 
                        \begin{pmatrix}
                            C_{0} & \cdots & C_{-n}\\
                            \vdots & \ddots & \vdots\\
                            C_{n} & \cdots & C_{0}
                        \end{pmatrix},
                        \quad
                        C_j =
                        \begin{pmatrix}
                            c_{j0} & \cdots & c_{j,-m}\\
                            \vdots & \ddots & \vdots \\
                            c_{j,m} & \cdots & c_{j0}
                        \end{pmatrix},
                    \end{equation}
                for $j \in \{-n, \cdots, n\}$, and where $c_{-k, -l} = c_{kl}^*$. Note this matrix is block-Toeplitz as described previously, and has dimension $(n + 1)(m + 1)\times(n + 1)(m + 1)$. This strange looking construction is standard as discussed in \cite{mv_frt_04}, and is motivated by single-indexed versions appearing in the proof of positive extensions and Schur complements in single-variable settings.
            \end{definition}
            
        \subsubsection{Proof of Theorem~\ref{thm:major_m_qsp_properties}}
        
            The proof of Theorem~\ref{thm:major_m_qsp_properties} will proceed similarly to its single-variable analogue, chiefly by induction. First, we introduce a related lemma to be used in the backwards ($\Leftarrow$) direction of the proof. This lemma underlies the need for Conjecture~\ref{conj:m_qsp_factorability}.
            
            \begin{lemma} \label{lemma:polynomial_square_ambiguity}
                Let $p, q$ be single-variable Laurent polynomials in $\mathbb{C}[x, x^{-1}]$ which satisfy the relation
                    \begin{equation} \label{eq:laurent_square_relation}
                        |p(x)|^2 = |q(x)|^2,
                    \end{equation}
                where $|p(x)|^2 = p(x)p^{*}(x)$ is the modulus squared of $p$ assuming $x$ real, and analogously for $q$. Then $q$ must be equal to $p$ up to exactly (1) some overall phase $e^{i\phi}$ and (2) complex conjugation of some subset of its roots.
                \begin{proof}
                    Condition (1) follows from the invariance of Eq.~\ref{eq:laurent_square_relation} up to an overall phase. Assume without loss of generality, then, that the leading coefficients of $p$ and $q$ are identical. Then one can express $|p(x)|^2 = |q(x)|^2$ in terms of its decomposition according to the fundamental theorem of algebra (assuming $p, q$ have degree $n$)
                        \begin{equation}
                            |p(x)|^2 = |q(x)|^2 = c x^{-2n}\prod_{k = 0}^{n} (x - a_k)(x - a_k^*),
                        \end{equation}
                    where the $a_k \in \mathbb{C}$ are not necessarily distinct. We see that $p$ and $q$ can thus be chosen to preserve this relation up to any of the $2^n$ (possibly non-distinct) choices between $a_k$ and $a_k^*$ as a root. This is precisely condition (2), and is the only other freedom without knowing other properties of $p, q$.
                \end{proof}
            \end{lemma}
            
            We begin the proof of Theorem~\ref{thm:major_m_qsp_properties} in earnest now, showing both directions, the second of which will depend on Conjecture~\ref{conj:m_qsp_factorability}.
            
            ($\Rightarrow$) The forward direction is relatively easy, taking $P = e^{i\phi_0}$ and $Q = 0$ which clearly satisfy properties (1-4). We can show that properties (1-3), as the overall operator is always unitary, are preserved by induction. Assume that for some length-$(n -1)$ protocol the inductive hypothesis holds and the unitary has form
                \begin{equation}
                    U_{s, \Phi}(a, b)
                    = 
                    \begin{pmatrix}
                        P & Q \\
                        -Q^* & P^*
                    \end{pmatrix},
                \end{equation}
            where $P, Q$ satisfy (1-4) in the Laurent picture. Without loss of generality we can apply $A(a)e^{i\phi_n\sigma_z}$ to yield a new M-QSP protocol with $s^\prime = s \cup \{0\}$ and $\Phi^\prime = \Phi \cup \{\phi_n\}$; this unitary $U_{s^\prime, \Phi^\prime}$ has form,
                \begin{equation} \label{eq:forward_m_qsp_iterate}
                    \frac{1}{2}\begin{pmatrix}
                        e^{i\phi_n}\left[(a + a^{-1})P + (a - a^{-1})Q\right] & e^{-i\phi_n}\left[(a - a^{-1})P + (a + a^{-1})Q\right]\\
                        * & *
                    \end{pmatrix},
                \end{equation}
            whose elements are still Laurent polynomials with degrees that match the desired bounds, condition (1), parity under $(a, b) \mapsto (a^{-1}, b^{-1})$ based on those known for $P, Q$, condition (2), parity under $a \mapsto -a$ and $b \mapsto -b$ based on those known for $P, Q$, condition (3), and the determinant condition, condition (4). These conditions clearly also hold if the operator appended was $B(b)$.
            
            ($\Leftarrow$) The reverse direction of this theorem is more involved, but again can be inspired by the single-variable case and its proof in \cite{gslw_19}, relying on a few extra lemmas to make the jump to the multivariable setting. We also make use of Conjecture~\ref{conj:m_qsp_factorability}, and indicate clearly where this is done. First we consider the trivial case in which the degree of $P$ is zero. Due to the symmetries specified, this must mean $|P(a = b = 1)| = 1$ and thus $Q(a = b = 1) = 0$. A simple solution to this is $\Phi = \{\phi_0, \pi/2, -\pi/2, \cdots \pi/2, -\pi/2\}$ in $\mathbb{R}^{n + 1}$ and $s = 00\cdots0$ of length $n$, which satisfies these conditions. This is the base case of our induction.
            
            The key step in the inductive argument involves inspection of the determinant, condition (4) in Theorem~\ref{thm:major_m_qsp_properties}, namely that for all $a, b$ such that $|a| = |b| = 1$ the relation
                \begin{equation}
                    P(a, b)P^*(a^{-1}, b^{-1}) + Q(a, b)Q^*(a^{-1}, b^{-1}) = 1
                \end{equation}
            holds identically. Our goal is to determine whether the application of an iterate of the form $e^{-i\phi_n}A^{\dag}(a)^{s_n}B^{\dag}(b)^{1 - s_n}$ can reduce the degrees of the embedded polynomials for any choice of $\phi_n$ and $s_n$. Inspection of this equation yields something very similar to Eq.~\ref{eq:forward_m_qsp_iterate}, namely a map
                \begin{equation}
                    \begin{pmatrix}
                        P & Q\\
                        -Q^* & P^*
                    \end{pmatrix}
                    \mapsto
                    \begin{pmatrix}
                        P^{\prime} & Q^{\prime}\\
                        -Q^{\prime*} & P^{\prime*}
                    \end{pmatrix},
                \end{equation}
            where $P^\prime, Q^\prime$ (the embedded polynomials after `peeling` off an M-QSP iterate $A(a)$ or $B(b)$) are of smaller degree in $a$ if $s_n = 1$ and in $b$ if $s_n = 0$, and where the explicit form of this circuit is, without loss of generality choosing $s_n = 1$ for now
                \begin{equation} \label{eq:backward_m_qsp_iterate}
                    \frac{1}{2}
                    \begin{pmatrix}
                        (a + a^{-1})e^{-i\phi_n}P + (a - a^{-1})e^{i\phi_n}Q & -(a - a^{-1})e^{-i\phi_n}P - (a + a^{-1})e^{i\phi_n}Q\\
                        * & *
                    \end{pmatrix}.
                \end{equation}
            The condition under which the top-left and top-right embedded polynomials are of lower degree than $P, Q$ is precisely that the coefficients of the maximal degree in $a$ of $P, Q$ differ exactly by an overall phase. Concretely, lowering the degrees of the embedded polynomials in either $a$ or $b$ (corresponding to $s_n = 1, 0$ respectively) requires that either of the two pairs of polynomial coefficients of the highest degrees of $a, b$ appearing in $P, Q$ respectively, namely
                \begin{align}
                    &P_{d_A}(b) = \sum_{k = -d_B}^{d_B} P_{d_A, k} b^{k},
                    \quad
                    Q_{d_A}(b) = \sum_{k = -d_B}^{d_B} Q_{d_A, k} b^{k}\label{eq:conj_cond_1}\\
                    &P_{d_B}(a) = \sum_{k = -d_A}^{d_A} P_{k,d_B} a^{k},
                    \quad
                    Q_{d_B}(a) = \sum_{k = -d_A}^{d_A} Q_{k, d_B} a^{k},\label{eq:conj_cond_2}
                \end{align}
            relate by an overall phase. Note that we discuss only the coefficient of the largest positive degree $d_A, d_B$ of $a$ or $b$; by the symmetries of these trigonometric Laurent polynomials, the largest negative degree in either variable will also identically vanish if this condition is satisfied.
            
            This condition is precisely the statement of Conjecture~\ref{conj:m_qsp_factorability}. Note that this condition also satisfies the requirement that the degrees of the bottom left and bottom right embedded polynomials are decreased, due to their trivial relation (complex conjugation) to the top left and top right embedded polynomials in Eq.~\ref{eq:backward_m_qsp_iterate}. Moreover, because this condition is a relation of coefficients, and these coefficients uniquely determine the roots of the corresponding single-variable polynomial coefficients of $P_{d_A}(b), Q_{d_A}(b)$ and $P_{d_B}(a), Q_{d_B}(a)$, Conjecture~\ref{conj:m_qsp_factorability} is both necessary and sufficient.
            
            If either of the conditions discussed in Conjecture~\ref{conj:m_qsp_factorability} holds, then there exists some choice of $\phi_n$ and $s_n$ such that the resulting lower-degree polynomials $P^\prime, Q^\prime$ still satisfy conditions (1-3) (and vacuously 4) from the statement of Theorem~\ref{thm:major_m_qsp_properties}. Moreover, if Conjecture~\ref{conj:m_qsp_factorability} holds, then this unitary must itself satisfy the inductive hypothesis, and the same process can be repeated to successively lower the degree in either $a$ or $b$ until the base case is satisfied and the result is shown. Note that it is sufficient to be able to lower the degree in $a$ or $b$, as as soon as one is reduced to the single-variable setting, the standard QSP theorems kick in, satisfying the conjecture vacuously.
            
            \begin{remark}
                Note that this argument leads easily to a proof of Corollary~\ref{cor:m_qsp_phase_readoff}, namely that any $P, Q$ arising from a unitary matrix built according to Definition~\ref{def:m_qsp} can be used, without explicit information about their constitutive $s, \Phi$, to efficiently determine an equivalent $s^\prime, \Phi^\prime$ by the classical method given above. To show this one needs to prove that the cases in which $s_n = 0$ and $s_n = 1$ are both possible (namely when both pairs of equations in Equations~\ref{eq:conj_cond_1} and \ref{eq:conj_cond_2} are satisfied) correspond only to $\phi_{n-1} \in \{-\pi, 0, \pi\}$ for $\phi_n$ restricted to $[-\pi, \pi]$.
                
                To show this we assume access to a description of the polynomials constituting an M-QSP protocol which takes the form
                    \begin{equation} \label{eq:s_ambiguity}
                        U_{\Phi, s} = U_{\Phi^\prime, s} A(a) e^{i\phi\sigma_z}B(b),
                    \end{equation}
                without loss of generality choosing the $B(b)$ iterate to have been applied last, and for the final M-QSP phase to have been identically zero implicitly. If we show that the only case in which the degree of the polynomials embedded in $U_{\Phi, s}$ can be lowered in either the variable $a$ or $b$ is when $\phi \in \{-\pi, 0, \pi\}$ (and thus one can commute $A(a)$ and $B(b)$), then one can use the phase read-off procedure discussed in the main proof of Theorem~\ref{thm:major_m_qsp_properties}.
                
                Proving this is easy; we simply write out the relevant matrix elements of  Eq.~\ref{eq:s_ambiguity}
                    \begin{align}
                        P &= \frac{1}{4}P^\prime 
                        \left[
                        e^{i\phi}\left(\frac{1}{ab} + \frac{a}{b} + \frac{b}{a} + ab\right) 
                        + 
                        e^{-i\phi}\left(\frac{1}{ab} - \frac{a}{b} - \frac{b}{a} + ab\right)
                        \right]\nonumber\\
                        &+ 
                        \frac{1}{4}Q^\prime
                        \left[
                        e^{i\phi}\left(-\frac{1}{ab} + \frac{a}{b} - \frac{b}{a} + ab\right) 
                        + 
                        e^{-i\phi}\left(-\frac{1}{ab} - \frac{a}{b} + \frac{b}{a} + ab\right)
                        \right]\\
                        Q &= \frac{1}{4}P^\prime 
                        \left[
                        e^{i\phi}\left(-\frac{1}{ab} - \frac{a}{b} + \frac{b}{a} + ab\right) 
                        + 
                        e^{-i\phi}\left(-\frac{1}{ab} + \frac{a}{b} - \frac{b}{a} + ab\right)
                        \right]\nonumber\\
                        &+ 
                        \frac{1}{4}Q^\prime
                        \left[
                        e^{i\phi}\left(\frac{1}{ab} - \frac{a}{b} - \frac{b}{a} + ab\right) 
                        + 
                        e^{-i\phi}\left(\frac{1}{ab} + \frac{a}{b} + \frac{b}{a} + ab\right)
                        \right].
                    \end{align}
                One can then look at the conditions under which the leading order single-variable coefficients of these two polynomials are identical up to an overall phase, in which case one can pull off an M-QSP iterate. For peeling a $B(b)$ or $A(a)$ iterate from this circuit we require that both pairs of leading coefficients differ by an overall phase: we write out these two pairs of equations:
                    \begin{align}
                        &\frac{1}{4}\tilde{P}
                        \left[
                            \frac{1}{b}\left(e^{i\phi} - e^{-i\phi}\right) + b\left(e^{i\phi} + e^{-i\phi}\right)
                        \right] 
                        + 
                        \frac{1}{4}\tilde{Q}
                        \left[
                            \frac{1}{b}\left(e^{i\phi} - e^{-i\phi}\right) + b\left(e^{i\phi} + e^{-i\phi}\right)
                        \right],\label{eq:a_iter_1}\\
                        &\frac{1}{4}\tilde{P}
                        \left[
                            -\frac{1}{b}\left(e^{i\phi} - e^{-i\phi}\right) + b\left(e^{i\phi} + e^{-i\phi}\right)
                        \right] 
                        + \frac{1}{4}\tilde{Q}
                        \left[
                            -\frac{1}{b}\left(e^{i\phi} - e^{-i\phi}\right) + b\left(e^{i\phi} + e^{-i\phi}\right)
                        \right],\label{eq:a_iter_2}
                    \end{align}
                for peeling off an $A(b)$ iterate (where $\tilde{P}, \tilde{Q}$ are the single-variable, in $b$, coefficients of the maximal degree terms of $P, Q$ in $a$), and likewise for peeling off a $B(b)$ iterate
                    \begin{align}
                        &\frac{1}{4}\tilde{P}
                        \left[
                            \frac{1}{a}\left(e^{i\phi} - e^{-i\phi}\right) + a\left(e^{i\phi} + e^{-i\phi}\right)
                        \right] 
                        + 
                        \frac{1}{4}\tilde{Q}
                        \left[
                            -\frac{1}{a}\left(e^{i\phi} - e^{-i\phi}\right) + a\left(e^{i\phi} + e^{-i\phi}\right)
                        \right],\label{eq:b_iter_1}\\
                        &\frac{1}{4}\tilde{P}
                        \left[
                            \frac{1}{a}\left(e^{i\phi} - e^{-i\phi}\right) + a\left(e^{i\phi} + e^{-i\phi}\right)
                        \right] 
                        + \frac{1}{4}\tilde{Q}
                        \left[
                            -\frac{1}{a}\left(e^{i\phi} - e^{-i\phi}\right) + a\left(e^{i\phi} + e^{-i\phi}\right)
                        \right].\label{eq:b_iter_2}
                    \end{align}
                In the case we're interested in, both of these pairs of polynomials need to differ by only an overall phase. Evidently this holds for Equations~\ref{eq:b_iter_1} and \ref{eq:b_iter_2} because they are identical. For Equations~\ref{eq:a_iter_1} and \ref{eq:a_iter_2}, the positive and negative powers of $b$ between the pair of equations now have a relative minus sign; equating these requires only the simple condition
                    \begin{equation}
                        e^{i\phi} - e^{-i\phi} = 2i\sin\phi =  0,
                    \end{equation}
                which in turn means that $\phi \in \{-\pi, 0, \pi\}$ as was desired. Note that the solution seemingly implied by taking $e^{i\phi} + e^{-i\phi} = 0$ is not valid, as we assume the rightmost implicit QSP phase is zero, and thus the phase relation between the pairs of equations must be trivial. This completes the proof of Corollary~\ref{cor:m_qsp_phase_readoff}, and indicates that there is no ambiguity in reading off $s$ from Definition~\ref{def:m_qsp} if the corresponding unitary was required to come from a product of iterates.
            \end{remark}
            
            This completes the proof of Theorem~\ref{thm:major_m_qsp_properties} under the assumption of Conjecture~\ref{conj:m_qsp_factorability}. Note that resolution of this conjecture, or \emph{any} additional possibly non-necessary conditions under which the property defined in Conjecture~\ref{conj:m_qsp_factorability} holds \emph{across the inductive} step, will allow the same proof as above to proceed. Moreover, we are able to show, as stated in Corollary~\ref{cor:m_qsp_phase_readoff}, that there is no fundamental difficulty in reading off M-QSP phases under the assurance that the corresponding unitary was built according to the definition of M-QSP; this corollary even extends to an arbitrary number of variables, though the corresponding Conjecture~\ref{conj:m_qsp_factorability} would be even more difficult to show. Nevertheless, the proof of this theorem presents the succinctly stated, minimal gauntlets that any attempt at a theory of M-QSP must address.
        
        \subsubsection{Proof of Theorem~\ref{thm:partial_m_qsp_properties}}
        
            \begin{theorem} \label{thm:m_frt}
                The multivariable Fejér-Riesz theorem (Theorem 1.1.3, equivalently generalized in Theorem 3.3.1, of \cite{mv_frt_04}). Let that the multivariable trigonometric Laurent polynomial
                    \begin{equation}
                        f(z, w) = \sum_{k = -n}^{n}\sum_{l = -m}^{m} f_{kl} z^k w^l
                    \end{equation}
                is strictly positive for all $|z| = |w| = 1$. Then there exists a \emph{stable} (Def.~\ref{def:stable_polynomials}) multivariable polynomial $p(z, w)$ such that $f(z, w) = |p(z, w)|^2$ with the following form:
                    \begin{equation}
                        p(z, w) = \sum_{k = 0}^{n}\sum_{l = 0}^{m} p_{kl}z^k w^l,
                    \end{equation}
                where stability means $p(z, w) \neq 0$ for $|z|, |w| \leq 1$, if and only if $\Gamma$ built from Fourier coefficients $c_{kl} = (\hat{1}/f)(k, l)$ of the reciprocal of $f$ (i.e., the matrix in Eq.~\ref{eq:low_rank_condition}, which is a doubly-indexed Toeplitz matrix as defined in Eq.~\ref{eq:gamma_toeplitz}) satisfies the following condition: the $(n + 1)m\times (m + 1)n$ submatrix of $\Gamma$ obtained by removing scalar (i.e., among the overall) rows $1 + j(m + 1)$ for $j \in \{0, \cdots, n\}$ and scalar columns $1, 2, \cdots, m + 1$ has rank $mn$. For ease of reference we note that this low-rank condition is the same as Eq.~\ref{eq:low_rank_condition}, namely
                    \begin{equation}
                        \left[(c_{u - v})_{u, v \in \Lambda\setminus\{0, 0\}}\right]^{-1}_{\subalign{&\{1, 2, \cdots, m\}\times\{0\} \\ &\{0\}\times\{1, 2, \cdots, n - m\}}} = 0,
                    \end{equation}
                where we note that $u, v$ don't run over the entirety of $\Lambda = \{0,1,\cdots, n\}\times\{0,1,\cdots, m\}$ for this matrix, and thus the resultant matrix is of dimension $[(n + 1)(m + 1)-1]\times[(n + 1)(m + 1)-1]$. Additionally note that up to $p$ being stable, the determined $p$ is unique up to an overall phase. For concrete computations of this matrix, see examples in \cite{mv_frt_04}.
            \end{theorem}
        
            The proof of Theorem~\ref{thm:partial_m_qsp_properties} proceeds similarly to its corresponding single-variable version, save the addition of a much stronger condition on a particular matrix relating to the specified \emph{strictly positive} real-valued trigonometric polynomial one wishes to embed using M-QSP. We simply define the additional objects necessary and reduce the statement of Theorem~\ref{thm:partial_m_qsp_properties} to a theorem in algebraic geometry: Theorem~\ref{thm:m_frt}. Additionally we supply a brief interpretation of the methods used to prove Theorem~\ref{thm:m_frt} (the full exposition of which runs about thirty pages in \cite{mv_frt_04}).
            
            We begin by stating the desired result of the theorem, introducing relevant variable names.  Consider a unitary matrix of the following form
                    \begin{equation} \label{eq:m_qsp_decomposed_form}
                        \begin{pmatrix}
                            P + iR & Q + iS\\
                            -Q + iS & P - iR
                        \end{pmatrix},
                    \end{equation}
            where $P, Q, R, S$, polynomials in $a, b$ take real values on $\mathbb{T}^2$. Note that that the $P, Q$ here are distinct from those defined in Theorem~\ref{thm:major_m_qsp_properties}, but that all unitaries defined in Theorem~\ref{thm:major_m_qsp_properties} may be suitably decomposed into a unitary of the form given in Eq.~\ref{eq:m_qsp_decomposed_form} by splitting the real and imaginary parts of the embeded polynomials on $\mathbb{T}^2$. The remaining work is to note that the unitarity of the matrix in Eq.~\ref{eq:m_qsp_decomposed_form} requires that the following relation holds
                \begin{equation} \label{eq:m_qsp_det_factoring}
                    1 - (P^2 + Q^2) = (R + iS)(R - iS).
                \end{equation}
            Consequently, as it was in the single variable case in Theorem~\ref{thm:partial_qsp_properties}, the existence of a matrix completion (i.e., a corresponding $R, S$) for a choice of $P, Q$ depends on the ability to factor $1 - (P^2 + Q^2)$ into a single square. In the single variable case the Fejér-Riesz theorem permitted this whenever $1 - (P^2 + Q^2)$ was non-negative. In the multivariable case we require that this quantity is positive, plus an additional series of constraints discussed below.
            
            The application of the multivariable Fejér-Riesz theorem (Theorem~\ref{thm:m_frt}) is clear; if the $\Gamma$ matrix corresponding to the Fourier coefficients of the inverse of $F = 1 - (P^2 + Q^2)$ satisfy the desired low-rank condition, then the multivariable single-variable polynomial function $T$ such that $F = |T|^2$ can be split into its real-valued ($R$) and imaginary-valued ($S$) components on $\mathbb{T}^2$. Note that because these Fourier components are real, and $P, Q$ have definite parity, then the Fourier components of the inverse are real as well, and thus $R$ and $iS$ have, as coefficients, purely real values. This means that the parity of $R$ and $S$ (the latter of which has purely imaginary coefficients) must be definite and opposite under inversion symmetry. All that is left is to ensure that these polynomials have the proper parity under negation symmetry $(a , b) \mapsto (-a, -b)$. But this is true obviously because $1 - (P^2 + Q^2)$ consists of powers of $a, b$ which are either $0$ or $2$ modulo $4$, corresponding uniquely to definite parity (odd, even under negation of both variables) for $T$. The final condition of Theorem~\ref{thm:partial_m_qsp_properties} (unitarity) is trivially satisfied by our satisfaction of Eq~\ref{eq:m_qsp_det_factoring}.
            
            Finally note that while the Fejér-Riesz theorem specifies decomposition into polynomials (i.e., sums of monomials with non-negative degrees, not Laurent polynomials), the same shifting argument can be used as in the single-variable case, namely noting that the following two products are equivalent, and thus we can make the substitution $T \mapsto T^\prime$ without worry.
                \begin{align}
                    TT^* 
                    &= 
                    \left[\sum_{j = 0}^{2(n - s)}\sum_{k = 0}^{2s} t_{jk} a^j b^k\right]
                    \left[\sum_{j = 0}^{2(n - s)}\sum_{k = 0}^{2s} t^*_{jk} a^{-j} b^{-k}\right],\\
                    T^\prime T^{\prime *} 
                    &= 
                    \left[\sum_{j = -(n - s)}^{(n - s)}\sum_{k = -s}^{s} t_{jk} a^j b^k\right]
                    \left[\sum_{j = -(n - s)}^{(n - s)}\sum_{k = -s}^{s} t^*_{jk} a^{-j} b^{-k}\right].
                \end{align}
            This completes the proof of Theorem~\ref{thm:partial_m_qsp_properties}, which at its core is far simpler than the proof of Theorem~\ref{thm:major_m_qsp_properties} beyond its reliance on the powerful albeit specific Theorem~\ref{thm:m_frt}. Moreover, it does not depend on Conjecture~\ref{conj:m_qsp_factorability}. It is purely a theorem about matrix completions (or positive extensions), where the ability to determine M-QSP phases for such a completion depends on either the validity of Conjecture~\ref{conj:m_qsp_factorability}, or that one's choice of $P, Q$ such that $1 - (P^2 + Q^2) > 0$ is judicious and happens to have a \emph{stable factorization}, which this theorem will efficiently verify the existence of and compute. If such a factorization exists, then Corollary~\ref{cor:m_qsp_phase_readoff} permits easy read-off of the M-QSP phases.

\end{document}